\documentclass[aip,pop,reprint,numerical,twocolumn]{revtex4-1}

\usepackage{amsfonts}
\usepackage{amsmath}
\usepackage{bm}
\usepackage{graphicx}
\usepackage{xcolor}
\usepackage{color}
\usepackage{amssymb}
\usepackage{natbib}

\usepackage{changes}
\definechangesauthor[name={Luca}, color=orange]{LSV}
\definechangesauthor[name={Denise}, color=blue]{DP}
\definechangesauthor[name={Oreste}, color=green]{OP}
\definechangesauthor[name={Franco}, color=gray]{FV}
\definechangesauthor[name={Antonella} ,color=magenta]{AG}
\definechangesauthor[name={Bill},color=purple]{Bill}

\def \beq {\begin{equation}}
\def \eeq {\end{equation}}
\def \bea {\begin{eqnarray}}
\def \eea {\end{eqnarray}}
\def \bfig {\begin{figure}}
\def \efig {\end{figure}}

\def \lab {\label}

\def \fr {\frac}

\renewcommand{\thefootnote}{${}^\dag$}

\begin{document}

\title{Velocity-Space Cascade in Magnetized Plasmas: Numerical Simulations}

\author{O. Pezzi$^1$, S. Servidio$^1$, D. Perrone$^{2,3}$, F. Valentini$^1$, L. Sorriso-Valvo$^4$, A. Greco$^1$, W.H. Matthaeus$^5$ and P. Veltri}

\affiliation{ $^1$Dipartimento di Fisica, Universit\`a della Calabria, I-87036 Cosenza, Italy\\
$^2$Department of Physics, Imperial College London, London SW7 2AZ, United Kingdom.\\
$^3$European Space Agency, ESAC, Madrid, Spain. \\
$^4$Nanotec-CNR, Sede di Rende, I-87036 Rende, Italy.\\
$^5$Bartol Research Institute and Department of Physics and Astronomy, University of Delaware, Newark, DE 19716, USA.}

\begin{abstract}
Plasma turbulence is studied via direct numerical simulations in a two-dimensional spatial geometry. Using a hybrid Vlasov-Maxwell model, we investigate the possibility of a velocity-space cascade. A novel theory of space plasma turbulence has been recently proposed by Servidio {\it et al.} [PRL, {\bf 119}, 205101 (2017)], supported by a three-dimensional Hermite decomposition  applied to spacecraft measurements, showing that velocity space fluctuations of the ion velocity distribution follow a broad-band, power-law Hermite spectrum $P(m)$, where $m$ is the Hermite index. We numerically explore these mechanisms in a more magnetized regime. We find that (1) the plasma reveals spectral anisotropy in velocity space, due to the presence of an external magnetic field (analogous to spatial anisotropy of fluid and plasma turbulence); (2) the distribution of energy follows the prediction $P(m)\sim m^{-2}$, proposed in the above theoretical-observational work; and (3) the velocity-space activity is intermittent in space, being enhanced close to coherent structures such as the reconnecting current sheets produced by turbulence. These results may be relevant to the nonlinear dynamics   weakly-collisional plasma in a wide variety of circumstances.
\end{abstract}
\date{\today}

\maketitle

Plasma turbulence is a challenging problem, involving a variety of complex nonlinear  phenomena. In the classical picture of turbulence, both in ordinary fluids and collisional plasmas, whenever energy is injected into the system, a cross-scale transfer occurs, producing smaller scales and leading eventually to energy conversion and dissipation. This non-linear behavior transfers energy from macroscopic gradients into small plasma eddies, waves and magnetic structures. The story becomes even more challenging in weakly collisional plasmas ---systems far from local thermal (Maxwellian) equilibrium. The absence of an equilibrium attractor leaves the plasma state free to explore the dual spatial-velocity phase space \citep{Huang}. This dynamics is responsible for  strong deformations of the particle distribution function (DF), commonly classified as rings, beams, temperature anisotropy, velocity-space vortices, and so on \citep{KrallTrivelpiece73,marsch82,HellingerEA06,BurchEA16}. In this multi-dimensional space, energy can be transfered nonlinearly from physical space to velocity space, and vice-versa, leading finally to the dissipation of the available energy through collisions \citep{Huang,MikhailovskiiEA74,PezziEA16,Howes18}.

The connection between turbulence and velocity-space deformations remains a great challenge for both theoretical and computational approaches. This scenario has been envisioned since the seminal works by Landau \citep{Landau46}. Recently, the study of phase-space fluctuations has become a topic of renewed interest within the plasma community \citep{Tatsuno09,ParkerEA16,TenBargeEA13-sheets,CerriEA18}. Many important and useful suggestions on the possibility of a spatial-velocity cascade have been recently proposed, in the framework of reduced models of plasma turbulence such as drift-wave and gyrokinetics \citep{KanekarEA15,SchekochihinEA16}. These concepts, closely related to strongly magnetized laboratory plasmas \citep{Eltgroth74}, need to be revised and further explored in the framework of space plasmas, where magnetic fluctuations are, very often, of the order of the mean magnetic field \citep{MGoldstein95}. In these regimes, indeed, nonlinear Landau damping and ion-cyclotron resonances, as well as interactions with current layers and zero-frequency structures, might occur in a more complex way \citep{HowesEA17,Pezzi2017turbulence}.

\begin{figure*}    
\includegraphics[width=\textwidth]{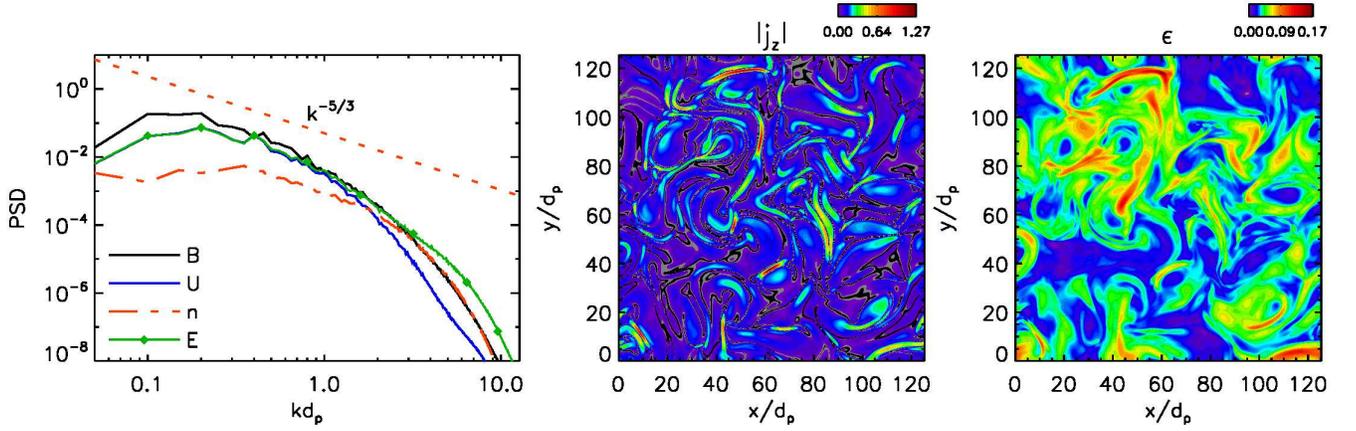}
\caption{Overview of the numerical results at the peak of the nonlinear activity, $t^*=49\Omega_{cp}^{-1}$. Left: omnidirectional perpendicular spectra of magnetic field $B$ (black), proton bulk speed $U$ (blue), electric field $E$ (green dotted) and proton density $n$ (red dashed), in code units. Center: contour plot of the current density $|j_z|$. Right: map exof the non-Maxwellianity function $\epsilon$, as defined in the text.}  
\label{fig:HVMSIM}
\end{figure*}

In the last decade or so, Vlasov-based simulations have been extensively used to investigate the complexity of plasma turbulence \citep{Howes08,ParasharEA09,servidio2012local,WanEA15,valentini2017transition,pezzi2017revisiting,ServidioEA16,FranciEA15a,FranciEA15,hellinger15}. Numerical experiments suggest a strong connection between turbulence and non-Maxwellian features in the particle DF \cite{GrecoEA12,ChasapisEA17,sorriso18local,sorriso2018statistical}. Recently, the unprecedented-resolution and high-accuracy of measurements from the Magnetospheric Multiscale Mission (MMS) \citep{BurchEA16} have enabled 
direct observation of the velocity space cascade in a space plasmas \citep{ServidioEA17}. In particular,  a three-dimensional (3D) Hermite decomposition has been applied to spacecraft measurements, showing that the ion velocity distribution has a broad-band Hermite spectrum $P(m)$, where $m$ is an Hermite mode index (see below). 
A Kolmogorov-like phenomenology has been proposed to interpret the observations, suggesting two types of phase-space cascade: (1) the isotropic cascade, with $P(m)\sim m^{-3/2}$, when the plasma is weakly magnetized (such as in the terrestrial, shocked magnetosheath), and (2) $P(m)\sim m^{-2}$, for more highly magnetized cases. Here we inspect the latter situations, exploring the possibility of an anisotropic cascade in velocity space, establishing its relation with spatial intermittency.

In this Letter we employ Hermite decomposition to analyze a hybrid Vlasov-Maxwell (HVM) simulation \cite{valentini2014hybrid,ServidioEA15} of collisonless plasma dynamics. 
Noise-free HVM simulations are well-suited for the study of the kinetic effects in turbulent collisionless plasmas.  We integrate the dimensionless HVM equations written as
\begin{eqnarray}
 &&\frac{\partial f}{\partial t} + {\bm v} \cdot \frac{\partial f}{\partial {\bm x} } +  \left ( {\bm E} + {\bm v} \times {\bm B} \right) \cdot \frac{\partial f}{\partial {\bm v}}=0, 
 \label{eq:HVMvlas} \\
&&\frac{\partial {\bf B}}{\partial t}\!=\! 
-{\bf \nabla}\!\times\!{\bm E}\!=\!{\bf \nabla}\!\times\!\left[ 
{\bm u}\times{\bm B}-\frac{{\bm j}\times{\bm B}}{n} + \frac{{\bf \nabla}P_e}{n} - \eta {\bm j}\right] 
 \label{eq:HVMfar}
\end{eqnarray}
where $f({\bm x}, {\bm v}, t)$ is the proton DF, ${\bm E}$ and ${\bm B}$ are the electric and magnetic fields, respectively. The current density is ${\bm j}={\bf \nabla}\times{\bm B}$, $n$ and $\bf u$ represent the first two moments of $f$, and $P_e$ is the isothermal pressure of the massless fluid electrons. In the above equations, time, velocities and lengths are respectively scaled to the inverse proton cyclotron frequency $\Omega_{cp}^{-1} = m_p c / e B_0$, to the  Alfv\'en speed $c_A = B_0 / \sqrt{4\pi n_{0} m_p}$, and to the proton skin depth $d_p= c_A/ \Omega_{cp}$, where $m_p$, $e$, $c$, $B_0$ and $n_{0}$ the proton mass, charge, the light speed, the background magnetic field and the equilibrium proton density. A small resistivity $\eta=2\times 10^{-2}$ is introduced to suppress numerical instabilities.

Equations~(\ref{eq:HVMvlas})--(\ref{eq:HVMfar}) are integrated in a 2.5D--3V phase space domain. We discretized the double-periodic spatial domain of size $L=2\pi\times 20 d_p$, with $N_x=N_y=512$ grid-points in each direction. The velocity domain is discretized by $N_{v_x}=N_{v_y}=N_{v_z}=71$ points in the range $v_{j}=\left[-5 v_{th},5v_{th}\right]\;(j=x,y,z)$ and boundary conditions impose $f(v_j>5 v_{th})=0$, being $v_{th}=\sqrt{k_{_B} T_{0}/m_p}$ the proton thermal speed, related to the Alfv\'en speed through $\beta=2 v_{th}^2/c_A^2=0.5$. Further details on the numerics can be found in Refs. \cite{valentini07,perrone2011role}.  The proton DF is initially Maxwellian, with uniform unit density.  In order to explore the possibility of a magnetized phase-space cascade, a uniform background out-of-plane magnetic field ${\bm B_0} = B_0 {\bm e}_z$ ($B_0=1$) is also imposed. The equilibrium is then perturbed through a $2D$ spectrum of Fourier modes, as described in \citet{ServidioEA15}. The r.m.s. level of fluctuations is $\delta B/B_0 = 1/3$. Note that this parameter range is very different from the MMS observations, where $\delta B/B_0\sim 2$ and $\beta\gg 1$ (weakly magnetized regime) \citep{ServidioEA17}.


We will discuss the results at the peak of the nonlinear activity, namely at $t^*=49\Omega_{cp}^{-1}$, where $\langle j_z^2\rangle$ reaches its maximum. To characterize the presence of small-scale fluctuations, the left panel of Fig. \ref{fig:HVMSIM} reports the omnidirectional perpendicular power spectral density of magnetic field (black), proton bulk speed (blue), electric field (green dotted) and proton density (red dashed), as a function of $k d_p$. The red dotted line indicates, as a reference, the Kolmogorov exponent $-5/3$. Magnetic fluctuations dominate at the large scales and the inertial range, while at smaller kinetic scales electric field spectral power is higher \citep{Bale05}. Moreover, although the large scales are essentially incompressible, at kinetic scales the compressibility increases \citep{NotePezziEA18}. 

Strong current sheets are evident in the shaded-contour of Fig. \ref{fig:HVMSIM} (central panel), which shows the out-of-plane current density $|j_z(x,y)|$ at $t=t^*$. As expected in turbulence, local narrow current layers develop and become important sites of reconnection and dissipation \citep{ServidioEA15,Retino:etal:2007,OsmanEA11,KarimabadiEA13}. Previous works have suggested that these intermittent regions are related to interesting non-Maxwellian features of the DF \citep{GrecoEA12}, a very well known property of magnetic reconnection \citep{Drake03,DrakeEA10,swisdak2016quantifying}. A simple non-Maxwellianity indicator\cite{GrecoEA12,pezzi2017colliding}, measuring deviations  of the particle DF from the corresponding Maxwellian $g$, has been defined as
\begin{equation}
 \epsilon({\bm x}) = \frac{1}{n}\sqrt{\int{(f-g)^2 d^3v}}.
 \label{eq:epsp}
\end{equation}
The right panel of Fig. \ref{fig:HVMSIM} shows $\epsilon(x,y)$ at $t=t^*$, suggesting that  the non-fluid activity is highly intermittent, correlated with the most intense current sheets. The scalar function $\epsilon$ quantifies the presence of high-order moments of the plasma, and includes moments of the proton DF, such as temperature anisotropy, heat flux, kurtosis and so on. It does not reveal, however, the particular structure of the velocity subspace. 

In order to quantify 
details of the phase-space cascade, we will adopt 
a 3D Hermite transform representation of $f({\bm x}, {\bm v})$, a valuable tool for plasma theory \citep{Grad49,ArmstrongMontgomery67,SchumerHolloway98}. A 1D basis can be defined as 
\beq
\psi_m(v)=\frac{ H_m\!\!\left(\!\fr{v-u}{v_{th}}\!\right) }{\sqrt{2^m m! \sqrt{\pi} v_{th} }} e^{ - \frac{(v-u)^2}{2 v_{th}^2}}, 
\lab{eq:psim}
\eeq
where $u$ and $v_{th}$ are now the local bulk and thermal speed, respectively, and $m\geq 0$ is an integer (we simplified the notation suppressing the spatial dependence). The eigenfunctions obey the orthogonality condition  $\int_{-\infty}^{\infty} \psi_m(v)\psi_l(v) d v = \delta_{m l}$.  Using this basis, one obtains a 3D representation  of the distribution function $f({\bm v}) = \sum_{\bm m} f_{\bm m} \psi_{\bm m}( {\bm v} )$. The above projection quantifies high-order corrections to the particle velocity DF, since the basis is shifted in the local fluid velocity frame, normalized to the ambient density and temperature. The projection in Eq.~(\ref{eq:psim}) is equivalent to shifting the Hermite grid in the local plasma frame, renormalizing such that the temperature is unity. Missing the above shift would generate a convolution with the central Maxwellian kernel and therefore a misleading spectrum. Moreover, a Gaussian quadrature is adopted \citep{Zhaohua14,ServidioEA17}, for efficiency, and to avoid spurious aliasing and convergence problems. We tested the accuracy of our Hermite transform, verifying that the Parseval-Plancerel spectral theorem is satisfied up to the machine precision.

\begin{figure}
\includegraphics[width=1\columnwidth]{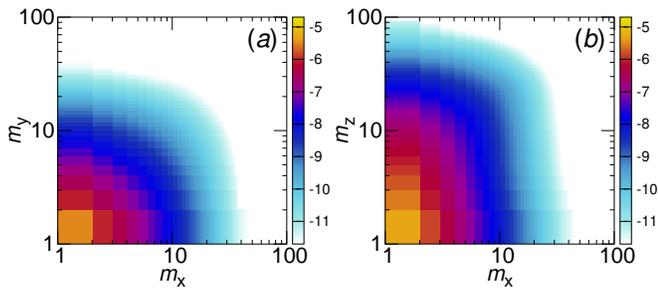}
\caption{Two-dimensional, reduced Hermite spectra, averaged over space. Panel (a) represents $P(m_x, m_y)$ (integrated over $m_z$), while (b) includes the anisotropy direction of the mean magnetic field $B_0$.}
\label{fig:2DSPHerm}
\end{figure}

Using the above procedure, the Hermite coefficients $f_{\bm m} = \int_{-\infty}^{\infty} f({\bm v}) \psi_{\bm m}( {\bm v} ) d^3 v$ have been computed. Note that the Hermite projection has a profound meaning for gases, since the index $m$ roughly corresponds to an order of the velocity moments \citep{SchekochihinEA16}:  the $m=1$ coefficient corresponds to bulk flow fluctuations; $m=2$ corresponds to temperature deformations; $m=3$ to heat flux perturbations, and so on.  Finally, it is worth noting that  an highly deformed $f({\bm v})$ would produce plasma enstrophy \citep{Knorr77}, defined as
\begin{equation}
\Omega({\bm x}) \equiv \int_{-\infty}^{\infty} \delta f^2({\bm x}, {\bm v}) d^3 v = \sum_{\bm m>0} \left[f_{\bm m}({\bm x})\right]^2, 
\label{eq:enst}
\end{equation}
where $\delta f$ indicates the difference from the ambient Maxwellian, as in Eq.~(\ref{eq:epsp}). It is interesting to note that the latter quantity is related to the Maxwellianity indicator $\epsilon$, $\Omega=\epsilon^2 n^2$, and is essentially the free energy in gyrokinetics \citep{SchekochihinEA16}.

\begin{figure}
\includegraphics[width=0.9\columnwidth]{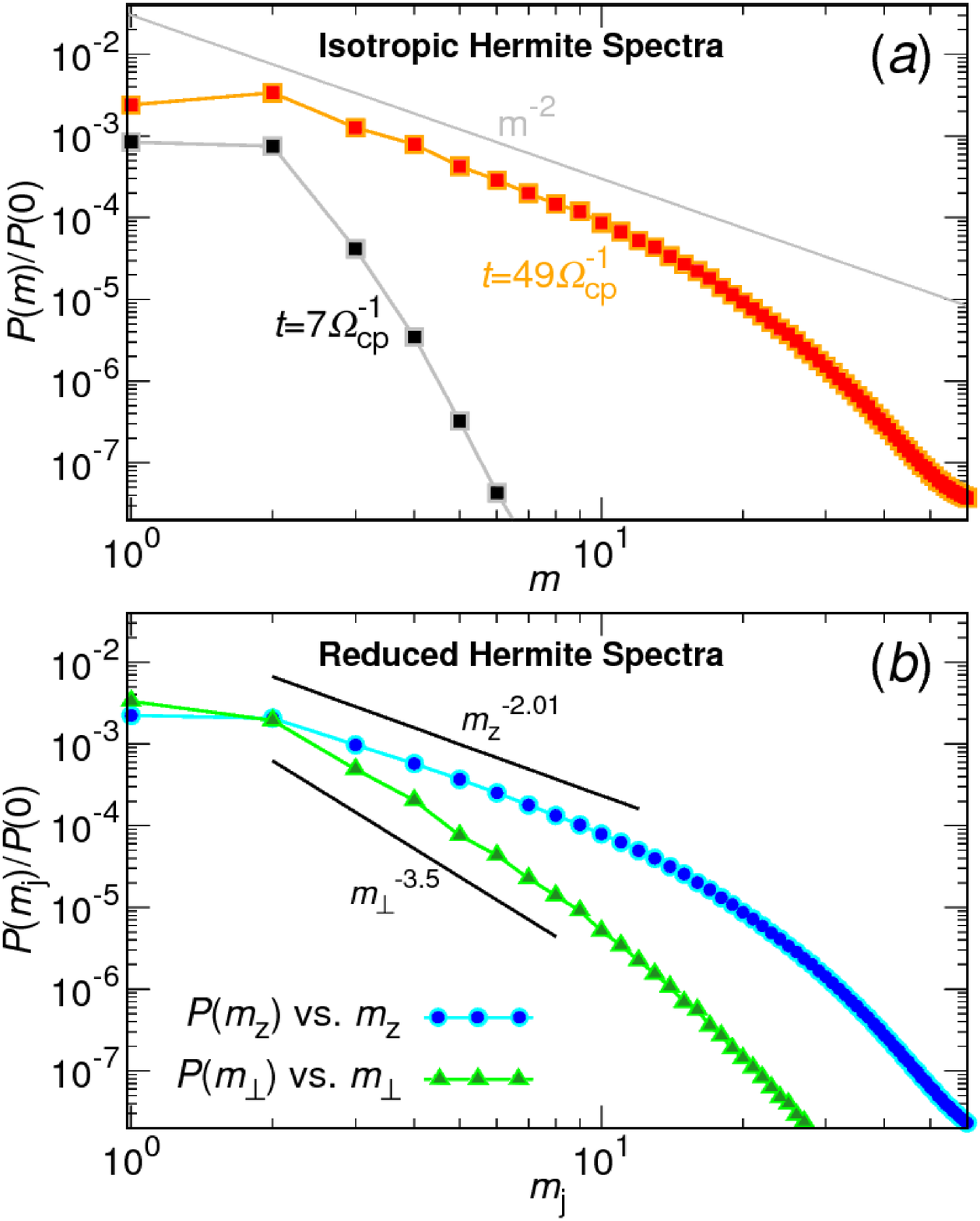}
\caption{
(a) Isotropic Hermite power spectrum, at two times of the simulation. The prediction for the magnetized case is reported as a reference. (b) Reduced spectra along and across the mean field, at $t=49 {\Omega}_{cp}^{-1}$. The power-law fit in the parallel direction is consistent with the prediction $\sim m_z^{-2}$.} 
\label{fig:isoSPHerm}
\end{figure}

In the Hermite transform, we set $N_m=100$ modes in each velocity direction, applying the projection to a subset of the original volume. In particular, we choose equally-spaced spatial points on a grid that is coarser than the original $512^2$, to reduce computational efforts (although the algorithm uses MPI parallel architecture).  We have checked that statistical convergence is attained for an ensemble of 32$\times$32 proton velocity DFs (not shown). In our analysis, we ensure convergence by using an ensemble of 64$\times$64 velocity DFs. From the coefficients $f_{\bm m}({\bm x})\equiv f_m(x,y,m_x,m_y,m_z)$, we define the enstrophy spectra as $P(m_x, m_y, m_z)= \langle f_{\bm m}({\bm x})^2\rangle$, where $\langle\dots\rangle$ indicates spatial average. 
Note that details of the phase space structure are lost 
when the Hermite spectrum is computed for 
poorly resolved  data, or when the 
spectrum is reduced (integrated over a velocity coordinate).





The 2D enstrophy spectrum is evaluated by reducing $P(m_x, m_y, m_z)$ in different directions, as for example $P(m_x, m_y) = \sum_{m_z=0}^{N_m} P(m_x, m_y, m_z)$. Figure \ref{fig:2DSPHerm} reports the 2D reduced spectra $P(m_x,m_y)$ and $P(m_x,m_z)$. 
The enstrophy is fairly isotropic in the plane $(m_x, m_y)$, perpendicular to the background magnetic field. On the other hand, an anisotropy is revealed when considering the direction of $B_0$, namely the $m_z$ axis: spectra are stretched in the parallel direction. This might indicate the presence of structures along the background field and the presence  of Landau resonances \citep{Kennel66}. Note also that this anisotropy was not recovered in the magnetosheath observations of MMS\cite{ServidioEA17}, since $\beta$ and $\delta B/B_0$ were much higher than in our simulation. This anisotropy is analogous to the spatial anisotropy commonly observed in plasmas, when a strong imposed field is present \citep{Shebalin83}. The velocity space anisotropy,  however, differs in that velocity gradients are stronger along the mean field and the cascade is inhibited across the mean field.

\begin{figure}
\includegraphics[width=1.01\columnwidth]{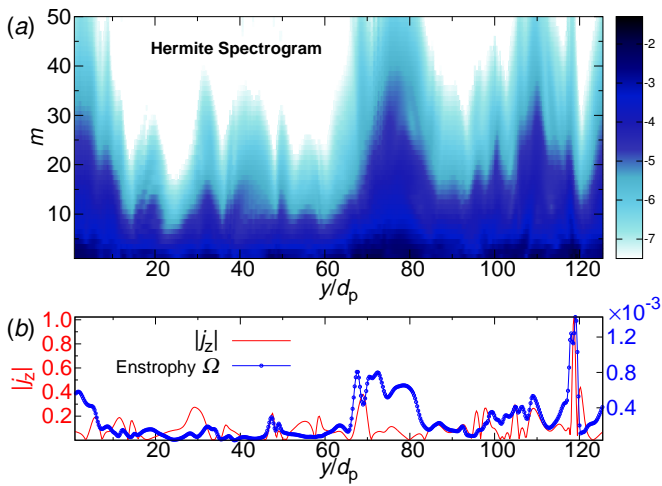}
\caption{(a) Hermite spectrogram, along a 1D cut through the simulation box, at $x^*\sim 50d_p$. The velocity space activity is highly intermittent. (b) Spatial profile at $x^*$ of the current density $|j_z|$, together with the plasma enstrophy defined by Eq. (\ref{eq:enst}). The velocity space cascade is correlated with intermittent coherent structures. } 
\label{fig:hermspe}
\end{figure}

We evaluated the isotropic (omnidirectional) 1D Hermite spectra, by summing $P(m_x,m_y,m_z)$ over concentric shells of unit width, i. e. $P(m) = \sum_{m-\frac12 < |{\bm m'}| \leq  m+\frac12} P({\bm m}') $. Figure \ref{fig:isoSPHerm}(a) shows the isotropic Hermite spectrum $P(m)/P(0)$ at $t=t^*=49\Omega_{cp}^{-1}$ (we normalized the spectrum to the mode $m=0$, which is the Maxwellian profile.) The spectrum shows a power-law behavior for the first decade, indicating the presence of phase-space cascade-like processes \cite{SircombeEA06,SchekochihinEA08,Tatsuno09,HatchEA14,KanekarEA15,ParkerDellar15,ParkerEA16,SchekochihinEA16,ServidioEA17}. The Hermite analysis on the HVM simulations shows a spectral break around $m\simeq 15$, where the artificial dissipation of the Eulerian scheme might affect the dynamics. In the same panel (a), we also plot the energy at an earlier time of the simulation, showing that the cascade has progressively emerged, as it would in physical space, by gradually filling in modes towards higher $m$-values, therefore creating finer velocity-space structures. In Fig.~(\ref{fig:isoSPHerm})-(b), we show the reduced spectra along the mean field (integrated over $m_x$ and $m_y$), and the isotropic perpendicular spectrum $P(m_\perp)$ (integrated over $m_z$ and over concentric perpendicular shells $m_\perp$). While $P(m_z)$ is consistent with the $m^{-2}$ phenomenological model, the reduced perpendicular spectrum $P(m_\perp)$ is much lower in energy and is steeper, with exponent close to $-3.5$. The significance of such anisotropy of the Hermite spectra will be investigated more in detail in future works.

In analogy with intermittency in turbulence, it is natural to ask whether or not the enstrophy transfer is homogeneous in space, as suggested by Fig. \ref{fig:HVMSIM} (center and right panels). To this aim, we define the {\it Hermite spectrogram} $P({\bm x}, m)$, the isotropic Hermite spectrum as a function of the position. This tool might be also useful for spacecraft measurements. Fig. \ref{fig:hermspe}-(a) shows $P({\bm x},m)$ along a one-dimensional spatial cut. The dual-space cascade is clearly intermittent: the spectrum amplitude and exponent depend on the position, with regions of low activity being interrupted by bursts of velocity-space activity. Only the ensemble average converges to the theoretical predictions represented in Fig. \ref{fig:isoSPHerm}. In panel (b), we show a spatial cut of the current density $|j_z|$ (red) and of the plasma enstrophy $\Omega$ (blue), suggesting that the velocity space cascade is correlated with the intermittent current structures.

Motivated by recent theories and observations\citep{ServidioEA17,Adkins18}, we have studied plasma turbulence via direct numerical simulations, in a simplified 2.5D-3V geometry. Using a hybrid Vlasov-Maxwell model, we observed that the proton velocity distribution function produces broad-band fluctuations in the $\bm v$--space. By using a 3D Hermite decomposition, we observed power-law Hermite spectra $P(m)$, indicative of a {\it velocity inertial range}. This velocity cascade establishes as the turbulence develops, resembling a mode-by-mode transfer, similar to the Kolmogorov phenomenology. Exploring a moderately magnetized case ($\delta B/B_0\sim1/3$ and $\beta=0.5$) we found that: (1) plasma manifests spectral anisotropy in velocity space, due to the presence of an external magnetic field (analogous to the {\it Shebalin effect}~\citep{Shebalin83}); (2) the distribution of energy follows the prediction $P(m)\sim m^{-2}$, and a much steeper exponent in the perpendicular direction, where $P(m_\perp)\sim m_\perp^{-3.5}$. Finally,  (3) the velocity space activity is intermittent in real space, and is enhanced close to coherent structures such as the reconnecting current layers produced by turbulence. In future works, we plan to explore different plasma regimes, as well as the role played by the dimensionality of the system and by the electron kinetics.  These results may be of fundamental significance as the space and astrophysical plasma communities move towards more complete 
understanding of the mechanisms leading to dissipation and  heating in turbulent plasmas.

\begin{acknowledgments}
This research was partially supported by AGS-1460130 (SHINE), NASA grants NNX14AI63G (Heliophysics Grand Challenge Theory), NNX15AB88G, NNX17AB79G, the Solar Probe Plus science team (ISOIS/Princeton subcontract SUB0000165), and by the MMS Theory and Modeling team, NNX14AC39G. FV and OP are supported by Agenzia Spaziale Italiana under contract ASI-INAF 2015-039-R.O. DP, SS ans LSV acknowledge support from the Faculty of the European Space Astronomy Centre (ESAC).  Computational support is provided by the Newton cluster at UNICAL. This work is partly supported by the International Space Science Institute (ISSI) in the framework of International Team 405 entitled “Current Sheets, Turbulence, Structures and Particle Acceleration in the Heliosphere”.
\end{acknowledgments}


\begin{thebibliography}{59}%
\makeatletter
\providecommand \@ifxundefined [1]{%
 \@ifx{#1\undefined}
}%
\providecommand \@ifnum [1]{%
 \ifnum #1\expandafter \@firstoftwo
 \else \expandafter \@secondoftwo
 \fi
}%
\providecommand \@ifx [1]{%
 \ifx #1\expandafter \@firstoftwo
 \else \expandafter \@secondoftwo
 \fi
}%
\providecommand \natexlab [1]{#1}%
\providecommand \enquote  [1]{``#1''}%
\providecommand \bibnamefont  [1]{#1}%
\providecommand \bibfnamefont [1]{#1}%
\providecommand \citenamefont [1]{#1}%
\providecommand \href@noop [0]{\@secondoftwo}%
\providecommand \href [0]{\begingroup \@sanitize@url \@href}%
\providecommand \@href[1]{\@@startlink{#1}\@@href}%
\providecommand \@@href[1]{\endgroup#1\@@endlink}%
\providecommand \@sanitize@url [0]{\catcode `\\12\catcode `\$12\catcode
  `\&12\catcode `\#12\catcode `\^12\catcode `\_12\catcode `\%12\relax}%
\providecommand \@@startlink[1]{}%
\providecommand \@@endlink[0]{}%
\providecommand \url  [0]{\begingroup\@sanitize@url \@url }%
\providecommand \@url [1]{\endgroup\@href {#1}{\urlprefix }}%
\providecommand \urlprefix  [0]{URL }%
\providecommand \Eprint [0]{\href }%
\providecommand \doibase [0]{http://dx.doi.org/}%
\providecommand \selectlanguage [0]{\@gobble}%
\providecommand \bibinfo  [0]{\@secondoftwo}%
\providecommand \bibfield  [0]{\@secondoftwo}%
\providecommand \translation [1]{[#1]}%
\providecommand \BibitemOpen [0]{}%
\providecommand \bibitemStop [0]{}%
\providecommand \bibitemNoStop [0]{.\EOS\space}%
\providecommand \EOS [0]{\spacefactor3000\relax}%
\providecommand \BibitemShut  [1]{\csname bibitem#1\endcsname}%
\let\auto@bib@innerbib\@empty
\bibitem [{\citenamefont {{Huang}}(1963)}]{Huang}%
  \BibitemOpen
  \bibfield  {author} {\bibinfo {author} {\bibfnamefont {K.}~\bibnamefont
  {{Huang}}},\ }\href@noop {} {\emph {\bibinfo {title} {Statistical Mechanics,
  New York: Wiley, 1963}}}\ (\bibinfo  {publisher} {~},\ \bibinfo {year}
  {1963})\BibitemShut {NoStop}%
\bibitem [{\citenamefont {{Krall}}\ and\ \citenamefont
  {{Trivelpiece}}(1973)}]{KrallTrivelpiece73}%
  \BibitemOpen
  \bibfield  {author} {\bibinfo {author} {\bibfnamefont {N.~A.}\ \bibnamefont
  {{Krall}}}\ and\ \bibinfo {author} {\bibfnamefont {A.~W.}\ \bibnamefont
  {{Trivelpiece}}},\ }\href@noop {} {\emph {\bibinfo {title} {Palaeogeography
  Palaeoclimatology Palaeoecology}}}\ (\bibinfo {year} {1973})\BibitemShut
  {NoStop}%
\bibitem [{\citenamefont {Marsch}\ \emph {et~al.}(1982)\citenamefont {Marsch},
  \citenamefont {M{\"u}hlh{\"a}user}, \citenamefont {Schwenn}, \citenamefont
  {Rosenbauer}, \citenamefont {Pilipp},\ and\ \citenamefont
  {Neubauer}}]{marsch82}%
  \BibitemOpen
  \bibfield  {author} {\bibinfo {author} {\bibfnamefont {E.}~\bibnamefont
  {Marsch}}, \bibinfo {author} {\bibfnamefont {K.-H.}\ \bibnamefont
  {M{\"u}hlh{\"a}user}}, \bibinfo {author} {\bibfnamefont {R.}~\bibnamefont
  {Schwenn}}, \bibinfo {author} {\bibfnamefont {H.}~\bibnamefont {Rosenbauer}},
  \bibinfo {author} {\bibfnamefont {W.}~\bibnamefont {Pilipp}}, \ and\ \bibinfo
  {author} {\bibfnamefont {F.}~\bibnamefont {Neubauer}},\ }\href@noop {}
  {\bibfield  {journal} {\bibinfo  {journal} {Journal of Geophysical Research:
  Space Physics (1978--2012)}\ }\textbf {\bibinfo {volume} {87}},\ \bibinfo
  {pages} {52} (\bibinfo {year} {1982})}\BibitemShut {NoStop}%
\bibitem [{\citenamefont {{Hellinger}}\ \emph {et~al.}(2006)\citenamefont
  {{Hellinger}}, \citenamefont {{Tr{\'a}vn{\'{\i}}{\v c}ek}}, \citenamefont
  {{Kasper}},\ and\ \citenamefont {{Lazarus}}}]{HellingerEA06}%
  \BibitemOpen
  \bibfield  {author} {\bibinfo {author} {\bibfnamefont {P.}~\bibnamefont
  {{Hellinger}}}, \bibinfo {author} {\bibfnamefont {P.}~\bibnamefont
  {{Tr{\'a}vn{\'{\i}}{\v c}ek}}}, \bibinfo {author} {\bibfnamefont {J.~C.}\
  \bibnamefont {{Kasper}}}, \ and\ \bibinfo {author} {\bibfnamefont {A.~J.}\
  \bibnamefont {{Lazarus}}},\ }\href {\doibase 10.1029/2006GL025925} {\bibfield
   {journal} {\bibinfo  {journal} {Geophys. Res. Lett.}\ }\textbf {\bibinfo
  {volume} {33}},\ \bibinfo {eid} {L09101} (\bibinfo {year}
  {2006})}\BibitemShut {NoStop}%
\bibitem [{\citenamefont {{Burch}}\ \emph {et~al.}(2016)\citenamefont
  {{Burch}}, \citenamefont {{Moore}}, \citenamefont {{Torbert}},\ and\
  \citenamefont {{Giles}}}]{BurchEA16}%
  \BibitemOpen
  \bibfield  {author} {\bibinfo {author} {\bibfnamefont {J.~L.}\ \bibnamefont
  {{Burch}}}, \bibinfo {author} {\bibfnamefont {T.~E.}\ \bibnamefont
  {{Moore}}}, \bibinfo {author} {\bibfnamefont {R.~B.}\ \bibnamefont
  {{Torbert}}}, \ and\ \bibinfo {author} {\bibfnamefont {B.~L.}\ \bibnamefont
  {{Giles}}},\ }\href {\doibase 10.1007/s11214-015-0164-9} {\bibfield
  {journal} {\bibinfo  {journal} {Space Sci. Rev.}\ }\textbf {\bibinfo {volume}
  {199}},\ \bibinfo {pages} {5} (\bibinfo {year} {2016})}\BibitemShut {NoStop}%
\bibitem [{\citenamefont {Mikhailovskii}(1974)}]{MikhailovskiiEA74}%
  \BibitemOpen
  \bibfield  {author} {\bibinfo {author} {\bibfnamefont {A.~B.}\ \bibnamefont
  {Mikhailovskii}},\ }\href@noop {} {\emph {\bibinfo {title} {{Theory of plasma
  instabilities. In: Instabilities in an Inhomogeneous Plasma, Vol. 2}}}}\
  (\bibinfo  {publisher} {{New York: Plenum}},\ \bibinfo {year}
  {1974})\BibitemShut {NoStop}%
\bibitem [{\citenamefont {{Pezzi}}, \citenamefont {{Valentini}},\ and\
  \citenamefont {{Veltri}}(2016)}]{PezziEA16}%
  \BibitemOpen
  \bibfield  {author} {\bibinfo {author} {\bibfnamefont {O.}~\bibnamefont
  {{Pezzi}}}, \bibinfo {author} {\bibfnamefont {F.}~\bibnamefont
  {{Valentini}}}, \ and\ \bibinfo {author} {\bibfnamefont {P.}~\bibnamefont
  {{Veltri}}},\ }\href {\doibase 10.1103/PhysRevLett.116.145001} {\bibfield
  {journal} {\bibinfo  {journal} {Physical Review Letters}\ }\textbf {\bibinfo
  {volume} {116}},\ \bibinfo {eid} {145001} (\bibinfo {year}
  {2016})}\BibitemShut {NoStop}%
\bibitem [{\citenamefont {{Howes}}(2018)}]{Howes18}%
  \BibitemOpen
  \bibfield  {author} {\bibinfo {author} {\bibfnamefont {G.~G.}\ \bibnamefont
  {{Howes}}},\ }\href@noop {} {\bibfield  {journal} {\bibinfo  {journal} {ArXiv
  e-prints}\ } (\bibinfo {year} {2018})},\ \Eprint
  {http://arxiv.org/abs/1802.04154} {arXiv:1802.04154 [physics.space-ph]}
  \BibitemShut {NoStop}%
\bibitem [{\citenamefont {{Landau}}(1946)}]{Landau46}%
  \BibitemOpen
  \bibfield  {author} {\bibinfo {author} {\bibfnamefont {L.}~\bibnamefont
  {{Landau}}},\ }\href@noop {} {\bibfield  {journal} {\bibinfo  {journal} {Zh.
  Eksp. Teor. Fiz.}\ }\textbf {\bibinfo {volume} {16}},\ \bibinfo {pages} {574}
  (\bibinfo {year} {1946})}\BibitemShut {NoStop}%
\bibitem [{\citenamefont {{Tatsuno}}\ \emph {et~al.}(2009)\citenamefont
  {{Tatsuno}}, \citenamefont {{Dorland}}, \citenamefont {{Schekochihin}},
  \citenamefont {{Plunk}}, \citenamefont {{Barnes}}, \citenamefont {{Cowley}},\
  and\ \citenamefont {{Howes}}}]{Tatsuno09}%
  \BibitemOpen
  \bibfield  {author} {\bibinfo {author} {\bibfnamefont {T.}~\bibnamefont
  {{Tatsuno}}}, \bibinfo {author} {\bibfnamefont {W.}~\bibnamefont
  {{Dorland}}}, \bibinfo {author} {\bibfnamefont {A.~A.}\ \bibnamefont
  {{Schekochihin}}}, \bibinfo {author} {\bibfnamefont {G.~G.}\ \bibnamefont
  {{Plunk}}}, \bibinfo {author} {\bibfnamefont {M.}~\bibnamefont {{Barnes}}},
  \bibinfo {author} {\bibfnamefont {S.~C.}\ \bibnamefont {{Cowley}}}, \ and\
  \bibinfo {author} {\bibfnamefont {G.~G.}\ \bibnamefont {{Howes}}},\ }\href
  {\doibase 10.1103/PhysRevLett.103.015003} {\bibfield  {journal} {\bibinfo
  {journal} {Physical Review Letters}\ }\textbf {\bibinfo {volume} {103}},\
  \bibinfo {eid} {015003} (\bibinfo {year} {2009})},\ \Eprint
  {http://arxiv.org/abs/0811.2538} {arXiv:0811.2538 [physics.plasm-ph]}
  \BibitemShut {NoStop}%
\bibitem [{\citenamefont {Parker}\ \emph {et~al.}(2016)\citenamefont {Parker},
  \citenamefont {Highcock}, \citenamefont {Schekochihin},\ and\ \citenamefont
  {Dellar}}]{ParkerEA16}%
  \BibitemOpen
  \bibfield  {author} {\bibinfo {author} {\bibfnamefont {J.~T.}\ \bibnamefont
  {Parker}}, \bibinfo {author} {\bibfnamefont {E.~G.}\ \bibnamefont
  {Highcock}}, \bibinfo {author} {\bibfnamefont {A.~A.}\ \bibnamefont
  {Schekochihin}}, \ and\ \bibinfo {author} {\bibfnamefont {P.~J.}\
  \bibnamefont {Dellar}},\ }\href {\doibase 10.1063/1.4958954} {\bibfield
  {journal} {\bibinfo  {journal} {Physics of Plasmas}\ }\textbf {\bibinfo
  {volume} {23}},\ \bibinfo {pages} {070703} (\bibinfo {year} {2016})},\
  \Eprint {http://arxiv.org/abs/https://doi.org/10.1063/1.4958954}
  {https://doi.org/10.1063/1.4958954} \BibitemShut {NoStop}%
\bibitem [{\citenamefont {{TenBarge}}\ and\ \citenamefont
  {{Howes}}(2013)}]{TenBargeEA13-sheets}%
  \BibitemOpen
  \bibfield  {author} {\bibinfo {author} {\bibfnamefont {J.~M.}\ \bibnamefont
  {{TenBarge}}}\ and\ \bibinfo {author} {\bibfnamefont {G.~G.}\ \bibnamefont
  {{Howes}}},\ }\href {\doibase 10.1088/2041-8205/771/2/L27} {\bibfield
  {journal} {\bibinfo  {journal} {The Astrophys. J. Lett.}\ }\textbf {\bibinfo
  {volume} {771}},\ \bibinfo {eid} {L27} (\bibinfo {year} {2013})},\ \Eprint
  {http://arxiv.org/abs/1304.2958} {arXiv:1304.2958 [physics.plasm-ph]}
  \BibitemShut {NoStop}%
\bibitem [{\citenamefont {{Cerri}}, \citenamefont {{Kunz}},\ and\ \citenamefont
  {{Califano}}(2018)}]{CerriEA18}%
  \BibitemOpen
  \bibfield  {author} {\bibinfo {author} {\bibfnamefont {S.~S.}\ \bibnamefont
  {{Cerri}}}, \bibinfo {author} {\bibfnamefont {M.~W.}\ \bibnamefont {{Kunz}}},
  \ and\ \bibinfo {author} {\bibfnamefont {F.}~\bibnamefont {{Califano}}},\
  }\href@noop {} {\bibfield  {journal} {\bibinfo  {journal} {ArXiv e-prints}\ }
  (\bibinfo {year} {2018})},\ \Eprint {http://arxiv.org/abs/1802.06133}
  {arXiv:1802.06133 [physics.plasm-ph]} \BibitemShut {NoStop}%
\bibitem [{\citenamefont {{Kanekar}}\ \emph {et~al.}(2015)\citenamefont
  {{Kanekar}}, \citenamefont {{Schekochihin}}, \citenamefont {{Dorland}},\ and\
  \citenamefont {{Loureiro}}}]{KanekarEA15}%
  \BibitemOpen
  \bibfield  {author} {\bibinfo {author} {\bibfnamefont {A.}~\bibnamefont
  {{Kanekar}}}, \bibinfo {author} {\bibfnamefont {A.~A.}\ \bibnamefont
  {{Schekochihin}}}, \bibinfo {author} {\bibfnamefont {W.}~\bibnamefont
  {{Dorland}}}, \ and\ \bibinfo {author} {\bibfnamefont {N.~F.}\ \bibnamefont
  {{Loureiro}}},\ }\href {\doibase 10.1017/S0022377814000622} {\bibfield
  {journal} {\bibinfo  {journal} {Journal of Plasma Physics}\ }\textbf
  {\bibinfo {volume} {81}},\ \bibinfo {eid} {305810104} (\bibinfo {year}
  {2015})},\ \Eprint {http://arxiv.org/abs/1403.6257} {arXiv:1403.6257
  [physics.plasm-ph]} \BibitemShut {NoStop}%
\bibitem [{\citenamefont {{Schekochihin}}\ \emph {et~al.}(2016)\citenamefont
  {{Schekochihin}}, \citenamefont {{Parker}}, \citenamefont {{Highcock}},
  \citenamefont {{Dellar}}, \citenamefont {{Dorland}},\ and\ \citenamefont
  {{Hammett}}}]{SchekochihinEA16}%
  \BibitemOpen
  \bibfield  {author} {\bibinfo {author} {\bibfnamefont {A.~A.}\ \bibnamefont
  {{Schekochihin}}}, \bibinfo {author} {\bibfnamefont {J.~T.}\ \bibnamefont
  {{Parker}}}, \bibinfo {author} {\bibfnamefont {E.~G.}\ \bibnamefont
  {{Highcock}}}, \bibinfo {author} {\bibfnamefont {P.~J.}\ \bibnamefont
  {{Dellar}}}, \bibinfo {author} {\bibfnamefont {W.}~\bibnamefont {{Dorland}}},
  \ and\ \bibinfo {author} {\bibfnamefont {G.~W.}\ \bibnamefont {{Hammett}}},\
  }\href {\doibase 10.1017/S0022377816000374} {\bibfield  {journal} {\bibinfo
  {journal} {Journal of Plasma Physics}\ }\textbf {\bibinfo {volume} {82}},\
  \bibinfo {eid} {905820212} (\bibinfo {year} {2016})},\ \Eprint
  {http://arxiv.org/abs/1508.05988} {arXiv:1508.05988 [physics.plasm-ph]}
  \BibitemShut {NoStop}%
\bibitem [{\citenamefont {{Eltgroth}}(1974)}]{Eltgroth74}%
  \BibitemOpen
  \bibfield  {author} {\bibinfo {author} {\bibfnamefont {P.~G.}\ \bibnamefont
  {{Eltgroth}}},\ }\href {\doibase 10.1063/1.1694939} {\bibfield  {journal}
  {\bibinfo  {journal} {Physics of Fluids}\ }\textbf {\bibinfo {volume} {17}},\
  \bibinfo {pages} {1602} (\bibinfo {year} {1974})}\BibitemShut {NoStop}%
\bibitem [{\citenamefont {Goldstein}, \citenamefont {Roberts},\ and\
  \citenamefont {Matthaeus}(1995)}]{MGoldstein95}%
  \BibitemOpen
  \bibfield  {author} {\bibinfo {author} {\bibfnamefont {M.~L.}\ \bibnamefont
  {Goldstein}}, \bibinfo {author} {\bibfnamefont {D.~A.}\ \bibnamefont
  {Roberts}}, \ and\ \bibinfo {author} {\bibfnamefont {W.~H.}\ \bibnamefont
  {Matthaeus}},\ }\href@noop {} {\bibfield  {journal} {\bibinfo  {journal}
  {Ann. Rev. Astron. Astrophys.}\ }\textbf {\bibinfo {volume} {33}},\ \bibinfo
  {pages} {283} (\bibinfo {year} {1995})}\BibitemShut {NoStop}%
\bibitem [{\citenamefont {{Howes}}, \citenamefont {{Klein}},\ and\
  \citenamefont {{Li}}(2017)}]{HowesEA17}%
  \BibitemOpen
  \bibfield  {author} {\bibinfo {author} {\bibfnamefont {G.~G.}\ \bibnamefont
  {{Howes}}}, \bibinfo {author} {\bibfnamefont {K.~G.}\ \bibnamefont
  {{Klein}}}, \ and\ \bibinfo {author} {\bibfnamefont {T.~C.}\ \bibnamefont
  {{Li}}},\ }\href {\doibase 10.1017/S0022377816001197} {\bibfield  {journal}
  {\bibinfo  {journal} {Journal of Plasma Physics}\ }\textbf {\bibinfo {volume}
  {83}},\ \bibinfo {eid} {705830102} (\bibinfo {year} {2017})}\BibitemShut
  {NoStop}%
\bibitem [{\citenamefont {Pezzi}\ \emph
  {et~al.}(2017{\natexlab{a}})\citenamefont {Pezzi}, \citenamefont {Malara},
  \citenamefont {Servidio}, \citenamefont {Valentini}, \citenamefont
  {Parashar}, \citenamefont {Matthaeus},\ and\ \citenamefont
  {Veltri}}]{Pezzi2017turbulence}%
  \BibitemOpen
  \bibfield  {author} {\bibinfo {author} {\bibfnamefont {O.}~\bibnamefont
  {Pezzi}}, \bibinfo {author} {\bibfnamefont {F.}~\bibnamefont {Malara}},
  \bibinfo {author} {\bibfnamefont {S.}~\bibnamefont {Servidio}}, \bibinfo
  {author} {\bibfnamefont {F.}~\bibnamefont {Valentini}}, \bibinfo {author}
  {\bibfnamefont {T.~N.}\ \bibnamefont {Parashar}}, \bibinfo {author}
  {\bibfnamefont {W.~H.}\ \bibnamefont {Matthaeus}}, \ and\ \bibinfo {author}
  {\bibfnamefont {P.}~\bibnamefont {Veltri}},\ }\href {\doibase
  10.1103/PhysRevE.96.023201} {\bibfield  {journal} {\bibinfo  {journal} {Phys.
  Rev. E}\ }\textbf {\bibinfo {volume} {96}},\ \bibinfo {pages} {023201}
  (\bibinfo {year} {2017}{\natexlab{a}})}\BibitemShut {NoStop}%
\bibitem [{\citenamefont {Howes}\ \emph {et~al.}(2008)\citenamefont {Howes},
  \citenamefont {Cowley}, \citenamefont {Dorland}, \citenamefont {Hammett},
  \citenamefont {Quataert},\ and\ \citenamefont {Schekochihin}}]{Howes08}%
  \BibitemOpen
  \bibfield  {author} {\bibinfo {author} {\bibfnamefont {G.~G.}\ \bibnamefont
  {Howes}}, \bibinfo {author} {\bibfnamefont {S.~C.}\ \bibnamefont {Cowley}},
  \bibinfo {author} {\bibfnamefont {W.}~\bibnamefont {Dorland}}, \bibinfo
  {author} {\bibfnamefont {G.~W.}\ \bibnamefont {Hammett}}, \bibinfo {author}
  {\bibfnamefont {E.}~\bibnamefont {Quataert}}, \ and\ \bibinfo {author}
  {\bibfnamefont {A.~A.}\ \bibnamefont {Schekochihin}},\ }\href {\doibase
  10.1029/2007JA012665} {\bibfield  {journal} {\bibinfo  {journal} {J. Geophys.
  Res.}\ }\textbf {\bibinfo {volume} {113}},\ \bibinfo {eid} {A05103} (\bibinfo
  {year} {2008}),\ 10.1029/2007JA012665}\BibitemShut {NoStop}%
\bibitem [{\citenamefont {{Parashar}}\ \emph {et~al.}(2009)\citenamefont
  {{Parashar}}, \citenamefont {{Shay}}, \citenamefont {{Cassak}},\ and\
  \citenamefont {{Matthaeus}}}]{ParasharEA09}%
  \BibitemOpen
  \bibfield  {author} {\bibinfo {author} {\bibfnamefont {T.~N.}\ \bibnamefont
  {{Parashar}}}, \bibinfo {author} {\bibfnamefont {M.~A.}\ \bibnamefont
  {{Shay}}}, \bibinfo {author} {\bibfnamefont {P.~A.}\ \bibnamefont
  {{Cassak}}}, \ and\ \bibinfo {author} {\bibfnamefont {W.~H.}\ \bibnamefont
  {{Matthaeus}}},\ }\href {\doibase 10.1063/1.3094062} {\bibfield  {journal}
  {\bibinfo  {journal} {Physics of Plasmas}\ }\textbf {\bibinfo {volume}
  {16}},\ \bibinfo {eid} {032310} (\bibinfo {year} {2009})}\BibitemShut
  {NoStop}%
\bibitem [{\citenamefont {Servidio}\ \emph {et~al.}(2012)\citenamefont
  {Servidio}, \citenamefont {Valentini}, \citenamefont {Califano},\ and\
  \citenamefont {Veltri}}]{servidio2012local}%
  \BibitemOpen
  \bibfield  {author} {\bibinfo {author} {\bibfnamefont {S.}~\bibnamefont
  {Servidio}}, \bibinfo {author} {\bibfnamefont {F.}~\bibnamefont {Valentini}},
  \bibinfo {author} {\bibfnamefont {F.}~\bibnamefont {Califano}}, \ and\
  \bibinfo {author} {\bibfnamefont {P.}~\bibnamefont {Veltri}},\ }\href@noop {}
  {\bibfield  {journal} {\bibinfo  {journal} {Physical review letters}\
  }\textbf {\bibinfo {volume} {108}},\ \bibinfo {pages} {045001} (\bibinfo
  {year} {2012})}\BibitemShut {NoStop}%
\bibitem [{\citenamefont {{Wan}}\ \emph {et~al.}(2015)\citenamefont {{Wan}},
  \citenamefont {{Matthaeus}}, \citenamefont {{Roytershteyn}}, \citenamefont
  {{Karimabadi}}, \citenamefont {{Parashar}}, \citenamefont {{Wu}},\ and\
  \citenamefont {{Shay}}}]{WanEA15}%
  \BibitemOpen
  \bibfield  {author} {\bibinfo {author} {\bibfnamefont {M.}~\bibnamefont
  {{Wan}}}, \bibinfo {author} {\bibfnamefont {W.~H.}\ \bibnamefont
  {{Matthaeus}}}, \bibinfo {author} {\bibfnamefont {V.}~\bibnamefont
  {{Roytershteyn}}}, \bibinfo {author} {\bibfnamefont {H.}~\bibnamefont
  {{Karimabadi}}}, \bibinfo {author} {\bibfnamefont {T.}~\bibnamefont
  {{Parashar}}}, \bibinfo {author} {\bibfnamefont {P.}~\bibnamefont {{Wu}}}, \
  and\ \bibinfo {author} {\bibfnamefont {M.}~\bibnamefont {{Shay}}},\ }\href
  {\doibase 10.1103/PhysRevLett.114.175002} {\bibfield  {journal} {\bibinfo
  {journal} {Physical Review Letters}\ }\textbf {\bibinfo {volume} {114}},\
  \bibinfo {eid} {175002} (\bibinfo {year} {2015})}\BibitemShut {NoStop}%
\bibitem [{\citenamefont {{Valentini, F.}}\ \emph {et~al.}(2017)\citenamefont
  {{Valentini, F.}}, \citenamefont {{V\'asconez, C. L.}}, \citenamefont
  {{Pezzi, O.}}, \citenamefont {{Servidio, S.}}, \citenamefont {{Malara, F.}},\
  and\ \citenamefont {{Pucci, F.}}}]{valentini2017transition}%
  \BibitemOpen
  \bibfield  {author} {\bibinfo {author} {\bibnamefont {{Valentini, F.}}},
  \bibinfo {author} {\bibnamefont {{V\'asconez, C. L.}}}, \bibinfo {author}
  {\bibnamefont {{Pezzi, O.}}}, \bibinfo {author} {\bibnamefont {{Servidio,
  S.}}}, \bibinfo {author} {\bibnamefont {{Malara, F.}}}, \ and\ \bibinfo
  {author} {\bibnamefont {{Pucci, F.}}},\ }\href {\doibase
  "10.1051/0004-6361/201629240"} {\bibfield  {journal} {\bibinfo  {journal}
  {"Astronomy and Astrophysics"}\ }\textbf {\bibinfo {volume} {599}},\ \bibinfo
  {pages} {"A8"} (\bibinfo {year} {2017})}\BibitemShut {NoStop}%
\bibitem [{\citenamefont {Pezzi}\ \emph
  {et~al.}(2017{\natexlab{b}})\citenamefont {Pezzi}, \citenamefont {Parashar},
  \citenamefont {Servidio}, \citenamefont {Valentini}, \citenamefont
  {V\'asconez}, \citenamefont {Yang}, \citenamefont {Malara}, \citenamefont
  {Matthaeus},\ and\ \citenamefont {Veltri}}]{pezzi2017revisiting}%
  \BibitemOpen
  \bibfield  {author} {\bibinfo {author} {\bibfnamefont {O.}~\bibnamefont
  {Pezzi}}, \bibinfo {author} {\bibfnamefont {T.~N.}\ \bibnamefont {Parashar}},
  \bibinfo {author} {\bibfnamefont {S.}~\bibnamefont {Servidio}}, \bibinfo
  {author} {\bibfnamefont {F.}~\bibnamefont {Valentini}}, \bibinfo {author}
  {\bibfnamefont {C.~L.}\ \bibnamefont {V\'asconez}}, \bibinfo {author}
  {\bibfnamefont {Y.}~\bibnamefont {Yang}}, \bibinfo {author} {\bibfnamefont
  {F.}~\bibnamefont {Malara}}, \bibinfo {author} {\bibfnamefont {W.~H.}\
  \bibnamefont {Matthaeus}}, \ and\ \bibinfo {author} {\bibfnamefont
  {P.}~\bibnamefont {Veltri}},\ }\href
  {http://stacks.iop.org/0004-637X/834/i=2/a=166} {\bibfield  {journal}
  {\bibinfo  {journal} {The Astrophysical Journal}\ }\textbf {\bibinfo {volume}
  {834}},\ \bibinfo {pages} {166} (\bibinfo {year}
  {2017}{\natexlab{b}})}\BibitemShut {NoStop}%
\bibitem [{\citenamefont {{Servidio}}\ \emph {et~al.}(2016)\citenamefont
  {{Servidio}}, \citenamefont {{Haynes}}, \citenamefont {{Matthaeus}},
  \citenamefont {{Burgess}}, \citenamefont {{Carbone}},\ and\ \citenamefont
  {{Veltri}}}]{ServidioEA16}%
  \BibitemOpen
  \bibfield  {author} {\bibinfo {author} {\bibfnamefont {S.}~\bibnamefont
  {{Servidio}}}, \bibinfo {author} {\bibfnamefont {C.~T.}\ \bibnamefont
  {{Haynes}}}, \bibinfo {author} {\bibfnamefont {W.~H.}\ \bibnamefont
  {{Matthaeus}}}, \bibinfo {author} {\bibfnamefont {D.}~\bibnamefont
  {{Burgess}}}, \bibinfo {author} {\bibfnamefont {V.}~\bibnamefont
  {{Carbone}}}, \ and\ \bibinfo {author} {\bibfnamefont {P.}~\bibnamefont
  {{Veltri}}},\ }\href {\doibase 10.1103/PhysRevLett.117.095101} {\bibfield
  {journal} {\bibinfo  {journal} {Physical Review Letters}\ }\textbf {\bibinfo
  {volume} {117}},\ \bibinfo {eid} {095101} (\bibinfo {year} {2016})},\ \Eprint
  {http://arxiv.org/abs/1608.01207} {arXiv:1608.01207 [physics.plasm-ph]}
  \BibitemShut {NoStop}%
\bibitem [{\citenamefont {{Franci}}\ \emph
  {et~al.}(2015{\natexlab{a}})\citenamefont {{Franci}}, \citenamefont
  {{Landi}}, \citenamefont {{Matteini}}, \citenamefont {{Verdini}},\ and\
  \citenamefont {{Hellinger}}}]{FranciEA15a}%
  \BibitemOpen
  \bibfield  {author} {\bibinfo {author} {\bibfnamefont {L.}~\bibnamefont
  {{Franci}}}, \bibinfo {author} {\bibfnamefont {S.}~\bibnamefont {{Landi}}},
  \bibinfo {author} {\bibfnamefont {L.}~\bibnamefont {{Matteini}}}, \bibinfo
  {author} {\bibfnamefont {A.}~\bibnamefont {{Verdini}}}, \ and\ \bibinfo
  {author} {\bibfnamefont {P.}~\bibnamefont {{Hellinger}}},\ }\href {\doibase
  10.1088/0004-637X/812/1/21} {\bibfield  {journal} {\bibinfo  {journal} {The
  Astrophysical Journal}\ }\textbf {\bibinfo {volume} {812}},\ \bibinfo {eid}
  {21} (\bibinfo {year} {2015}{\natexlab{a}})},\ \Eprint
  {http://arxiv.org/abs/1506.05999} {arXiv:1506.05999 [astro-ph.SR]}
  \BibitemShut {NoStop}%
\bibitem [{\citenamefont {{Franci}}\ \emph
  {et~al.}(2015{\natexlab{b}})\citenamefont {{Franci}}, \citenamefont
  {{Verdini}}, \citenamefont {{Matteini}}, \citenamefont {{Landi}},\ and\
  \citenamefont {{Hellinger}}}]{FranciEA15}%
  \BibitemOpen
  \bibfield  {author} {\bibinfo {author} {\bibfnamefont {L.}~\bibnamefont
  {{Franci}}}, \bibinfo {author} {\bibfnamefont {A.}~\bibnamefont {{Verdini}}},
  \bibinfo {author} {\bibfnamefont {L.}~\bibnamefont {{Matteini}}}, \bibinfo
  {author} {\bibfnamefont {S.}~\bibnamefont {{Landi}}}, \ and\ \bibinfo
  {author} {\bibfnamefont {P.}~\bibnamefont {{Hellinger}}},\ }\href {\doibase
  10.1088/2041-8205/804/2/L39} {\bibfield  {journal} {\bibinfo  {journal} {The
  Astrophys. J. Lett.}\ }\textbf {\bibinfo {volume} {804}},\ \bibinfo {eid}
  {L39} (\bibinfo {year} {2015}{\natexlab{b}})},\ \Eprint
  {http://arxiv.org/abs/1503.05457} {arXiv:1503.05457 [astro-ph.SR]}
  \BibitemShut {NoStop}%
\bibitem [{\citenamefont {{Hellinger}}\ \emph {et~al.}(2015)\citenamefont
  {{Hellinger}}, \citenamefont {{Matteini}}, \citenamefont {{Landi}},
  \citenamefont {{Verdini}}, \citenamefont {{Franci}},\ and\ \citenamefont
  {{Tr{\'a}vn{\'{\i}}{\v c}ek}}}]{hellinger15}%
  \BibitemOpen
  \bibfield  {author} {\bibinfo {author} {\bibfnamefont {P.}~\bibnamefont
  {{Hellinger}}}, \bibinfo {author} {\bibfnamefont {L.}~\bibnamefont
  {{Matteini}}}, \bibinfo {author} {\bibfnamefont {S.}~\bibnamefont {{Landi}}},
  \bibinfo {author} {\bibfnamefont {A.}~\bibnamefont {{Verdini}}}, \bibinfo
  {author} {\bibfnamefont {L.}~\bibnamefont {{Franci}}}, \ and\ \bibinfo
  {author} {\bibfnamefont {P.}~\bibnamefont {{Tr{\'a}vn{\'{\i}}{\v c}ek}}},\
  }\href {\doibase 10.1088/2041-8205/811/2/L32} {\bibfield  {journal} {\bibinfo
   {journal} {The Astrophysical Journal Letters}\ }\textbf {\bibinfo {volume}
  {811}},\ \bibinfo {eid} {L32} (\bibinfo {year} {2015})},\ \Eprint
  {http://arxiv.org/abs/1508.03159} {arXiv:1508.03159 [physics.space-ph]}
  \BibitemShut {NoStop}%
\bibitem [{\citenamefont {Greco}\ \emph {et~al.}(2012)\citenamefont {Greco},
  \citenamefont {Valentini}, \citenamefont {Servidio},\ and\ \citenamefont
  {Matthaeus}}]{GrecoEA12}%
  \BibitemOpen
  \bibfield  {author} {\bibinfo {author} {\bibfnamefont {A.}~\bibnamefont
  {Greco}}, \bibinfo {author} {\bibfnamefont {F.}~\bibnamefont {Valentini}},
  \bibinfo {author} {\bibfnamefont {S.}~\bibnamefont {Servidio}}, \ and\
  \bibinfo {author} {\bibfnamefont {W.}~\bibnamefont {Matthaeus}},\ }\href@noop
  {} {\bibfield  {journal} {\bibinfo  {journal} {Physical Review E}\ }\textbf
  {\bibinfo {volume} {86}},\ \bibinfo {pages} {066405} (\bibinfo {year}
  {2012})}\BibitemShut {NoStop}%
\bibitem [{\citenamefont {{Chasapis}}\ \emph {et~al.}(2017)\citenamefont
  {{Chasapis}}, \citenamefont {{Matthaeus}}, \citenamefont {{Parashar}},
  \citenamefont {{LeContel}}, \citenamefont {{Retin{\`o}}}, \citenamefont
  {{Breuillard}}, \citenamefont {{Khotyaintsev}}, \citenamefont {{Vaivads}},
  \citenamefont {{Lavraud}}, \citenamefont {{Eriksson}}, \citenamefont
  {{Moore}}, \citenamefont {{Burch}}, \citenamefont {{Torbert}}, \citenamefont
  {{Lindqvist}}, \citenamefont {{Ergun}}, \citenamefont {{Marklund}},
  \citenamefont {{Goodrich}}, \citenamefont {{Wilder}}, \citenamefont
  {{Chutter}}, \citenamefont {{Needell}}, \citenamefont {{Rau}}, \citenamefont
  {{Dors}}, \citenamefont {{Russell}}, \citenamefont {{Le}}, \citenamefont
  {{Magnes}}, \citenamefont {{Strangeway}}, \citenamefont {{Bromund}},
  \citenamefont {{Leinweber}}, \citenamefont {{Plaschke}}, \citenamefont
  {{Fischer}}, \citenamefont {{Anderson}}, \citenamefont {{Pollock}},
  \citenamefont {{Giles}}, \citenamefont {{Paterson}}, \citenamefont
  {{Dorelli}}, \citenamefont {{Gershman}}, \citenamefont {{Avanov}},\ and\
  \citenamefont {{Saito}}}]{ChasapisEA17}%
  \BibitemOpen
  \bibfield  {author} {\bibinfo {author} {\bibfnamefont {A.}~\bibnamefont
  {{Chasapis}}}, \bibinfo {author} {\bibfnamefont {W.~H.}\ \bibnamefont
  {{Matthaeus}}}, \bibinfo {author} {\bibfnamefont {T.~N.}\ \bibnamefont
  {{Parashar}}}, \bibinfo {author} {\bibfnamefont {O.}~\bibnamefont
  {{LeContel}}}, \bibinfo {author} {\bibfnamefont {A.}~\bibnamefont
  {{Retin{\`o}}}}, \bibinfo {author} {\bibfnamefont {H.}~\bibnamefont
  {{Breuillard}}}, \bibinfo {author} {\bibfnamefont {Y.}~\bibnamefont
  {{Khotyaintsev}}}, \bibinfo {author} {\bibfnamefont {A.}~\bibnamefont
  {{Vaivads}}}, \bibinfo {author} {\bibfnamefont {B.}~\bibnamefont
  {{Lavraud}}}, \bibinfo {author} {\bibfnamefont {E.}~\bibnamefont
  {{Eriksson}}}, \bibinfo {author} {\bibfnamefont {T.~E.}\ \bibnamefont
  {{Moore}}}, \bibinfo {author} {\bibfnamefont {J.~L.}\ \bibnamefont
  {{Burch}}}, \bibinfo {author} {\bibfnamefont {R.~B.}\ \bibnamefont
  {{Torbert}}}, \bibinfo {author} {\bibfnamefont {P.-A.}\ \bibnamefont
  {{Lindqvist}}}, \bibinfo {author} {\bibfnamefont {R.~E.}\ \bibnamefont
  {{Ergun}}}, \bibinfo {author} {\bibfnamefont {G.}~\bibnamefont {{Marklund}}},
  \bibinfo {author} {\bibfnamefont {K.~A.}\ \bibnamefont {{Goodrich}}},
  \bibinfo {author} {\bibfnamefont {F.~D.}\ \bibnamefont {{Wilder}}}, \bibinfo
  {author} {\bibfnamefont {M.}~\bibnamefont {{Chutter}}}, \bibinfo {author}
  {\bibfnamefont {J.}~\bibnamefont {{Needell}}}, \bibinfo {author}
  {\bibfnamefont {D.}~\bibnamefont {{Rau}}}, \bibinfo {author} {\bibfnamefont
  {I.}~\bibnamefont {{Dors}}}, \bibinfo {author} {\bibfnamefont {C.~T.}\
  \bibnamefont {{Russell}}}, \bibinfo {author} {\bibfnamefont {G.}~\bibnamefont
  {{Le}}}, \bibinfo {author} {\bibfnamefont {W.}~\bibnamefont {{Magnes}}},
  \bibinfo {author} {\bibfnamefont {R.~J.}\ \bibnamefont {{Strangeway}}},
  \bibinfo {author} {\bibfnamefont {K.~R.}\ \bibnamefont {{Bromund}}}, \bibinfo
  {author} {\bibfnamefont {H.~K.}\ \bibnamefont {{Leinweber}}}, \bibinfo
  {author} {\bibfnamefont {F.}~\bibnamefont {{Plaschke}}}, \bibinfo {author}
  {\bibfnamefont {D.}~\bibnamefont {{Fischer}}}, \bibinfo {author}
  {\bibfnamefont {B.~J.}\ \bibnamefont {{Anderson}}}, \bibinfo {author}
  {\bibfnamefont {C.~J.}\ \bibnamefont {{Pollock}}}, \bibinfo {author}
  {\bibfnamefont {B.~L.}\ \bibnamefont {{Giles}}}, \bibinfo {author}
  {\bibfnamefont {W.~R.}\ \bibnamefont {{Paterson}}}, \bibinfo {author}
  {\bibfnamefont {J.}~\bibnamefont {{Dorelli}}}, \bibinfo {author}
  {\bibfnamefont {D.~J.}\ \bibnamefont {{Gershman}}}, \bibinfo {author}
  {\bibfnamefont {L.}~\bibnamefont {{Avanov}}}, \ and\ \bibinfo {author}
  {\bibfnamefont {Y.}~\bibnamefont {{Saito}}},\ }\href {\doibase
  10.3847/1538-4357/836/2/247} {\bibfield  {journal} {\bibinfo  {journal} {The
  Astrophysical Journal}\ }\textbf {\bibinfo {volume} {836}},\ \bibinfo {eid}
  {247} (\bibinfo {year} {2017})}\BibitemShut {NoStop}%
\bibitem [{\citenamefont {Sorriso-Valvo}\ \emph {et~al.}()\citenamefont
  {Sorriso-Valvo}, \citenamefont {Perrone}, \citenamefont {Pezzi},
  \citenamefont {Valentini}, \citenamefont {Servidio}, \citenamefont
  {Zouganelis},\ and\ \citenamefont {Veltri}}]{sorriso18local}%
  \BibitemOpen
  \bibfield  {author} {\bibinfo {author} {\bibfnamefont {L.}~\bibnamefont
  {Sorriso-Valvo}}, \bibinfo {author} {\bibfnamefont {D.}~\bibnamefont
  {Perrone}}, \bibinfo {author} {\bibfnamefont {O.}~\bibnamefont {Pezzi}},
  \bibinfo {author} {\bibfnamefont {F.}~\bibnamefont {Valentini}}, \bibinfo
  {author} {\bibfnamefont {S.}~\bibnamefont {Servidio}}, \bibinfo {author}
  {\bibfnamefont {Y.}~\bibnamefont {Zouganelis}}, \ and\ \bibinfo {author}
  {\bibfnamefont {P.}~\bibnamefont {Veltri}},\ }\href@noop {} {\bibinfo
  {journal} {Journal of Plasma Physics}\ }\BibitemShut {NoStop}%
\bibitem [{\citenamefont {Sorriso-Valvo}\ \emph {et~al.}(2018)\citenamefont
  {Sorriso-Valvo}, \citenamefont {Carbone}, \citenamefont {Perri},
  \citenamefont {Greco}, \citenamefont {Marino},\ and\ \citenamefont
  {Bruno}}]{sorriso2018statistical}%
  \BibitemOpen
\bibfield  {journal} {  }\bibfield  {author} {\bibinfo {author} {\bibfnamefont
  {L.}~\bibnamefont {Sorriso-Valvo}}, \bibinfo {author} {\bibfnamefont
  {F.}~\bibnamefont {Carbone}}, \bibinfo {author} {\bibfnamefont
  {S.}~\bibnamefont {Perri}}, \bibinfo {author} {\bibfnamefont
  {A.}~\bibnamefont {Greco}}, \bibinfo {author} {\bibfnamefont
  {R.}~\bibnamefont {Marino}}, \ and\ \bibinfo {author} {\bibfnamefont
  {R.}~\bibnamefont {Bruno}},\ }\href {\doibase 10.1007/s11207-017-1229-6}
  {\bibfield  {journal} {\bibinfo  {journal} {Solar Physics}\ }\textbf
  {\bibinfo {volume} {293}},\ \bibinfo {pages} {10} (\bibinfo {year}
  {2018})}\BibitemShut {NoStop}%
\bibitem [{\citenamefont {Servidio}\ \emph {et~al.}(2017)\citenamefont
  {Servidio}, \citenamefont {Chasapis}, \citenamefont {Matthaeus},
  \citenamefont {Perrone}, \citenamefont {Valentini}, \citenamefont {Parashar},
  \citenamefont {Veltri}, \citenamefont {Gershman}, \citenamefont {Russell},
  \citenamefont {Giles}, \citenamefont {Fuselier}, \citenamefont {Phan},\ and\
  \citenamefont {Burch}}]{ServidioEA17}%
  \BibitemOpen
  \bibfield  {author} {\bibinfo {author} {\bibfnamefont {S.}~\bibnamefont
  {Servidio}}, \bibinfo {author} {\bibfnamefont {A.}~\bibnamefont {Chasapis}},
  \bibinfo {author} {\bibfnamefont {W.~H.}\ \bibnamefont {Matthaeus}}, \bibinfo
  {author} {\bibfnamefont {D.}~\bibnamefont {Perrone}}, \bibinfo {author}
  {\bibfnamefont {F.}~\bibnamefont {Valentini}}, \bibinfo {author}
  {\bibfnamefont {T.~N.}\ \bibnamefont {Parashar}}, \bibinfo {author}
  {\bibfnamefont {P.}~\bibnamefont {Veltri}}, \bibinfo {author} {\bibfnamefont
  {D.}~\bibnamefont {Gershman}}, \bibinfo {author} {\bibfnamefont {C.~T.}\
  \bibnamefont {Russell}}, \bibinfo {author} {\bibfnamefont {B.}~\bibnamefont
  {Giles}}, \bibinfo {author} {\bibfnamefont {S.~A.}\ \bibnamefont {Fuselier}},
  \bibinfo {author} {\bibfnamefont {T.~D.}\ \bibnamefont {Phan}}, \ and\
  \bibinfo {author} {\bibfnamefont {J.}~\bibnamefont {Burch}},\ }\href
  {\doibase 10.1103/PhysRevLett.119.205101} {\bibfield  {journal} {\bibinfo
  {journal} {Phys. Rev. Lett.}\ }\textbf {\bibinfo {volume} {119}},\ \bibinfo
  {pages} {205101} (\bibinfo {year} {2017})}\BibitemShut {NoStop}%
\bibitem [{\citenamefont {Valentini}\ \emph {et~al.}(2014)\citenamefont
  {Valentini}, \citenamefont {Servidio}, \citenamefont {Perrone}, \citenamefont
  {Califano}, \citenamefont {Matthaeus},\ and\ \citenamefont
  {Veltri}}]{valentini2014hybrid}%
  \BibitemOpen
  \bibfield  {author} {\bibinfo {author} {\bibfnamefont {F.}~\bibnamefont
  {Valentini}}, \bibinfo {author} {\bibfnamefont {S.}~\bibnamefont {Servidio}},
  \bibinfo {author} {\bibfnamefont {D.}~\bibnamefont {Perrone}}, \bibinfo
  {author} {\bibfnamefont {F.}~\bibnamefont {Califano}}, \bibinfo {author}
  {\bibfnamefont {W.}~\bibnamefont {Matthaeus}}, \ and\ \bibinfo {author}
  {\bibfnamefont {P.}~\bibnamefont {Veltri}},\ }\href@noop {} {\bibfield
  {journal} {\bibinfo  {journal} {Physics of Plasmas}\ }\textbf {\bibinfo
  {volume} {21}},\ \bibinfo {pages} {082307} (\bibinfo {year}
  {2014})}\BibitemShut {NoStop}%
\bibitem [{\citenamefont {{Servidio}}\ \emph {et~al.}(2015)\citenamefont
  {{Servidio}}, \citenamefont {{Valentini}}, \citenamefont {{Perrone}},
  \citenamefont {{Greco}}, \citenamefont {{Califano}}, \citenamefont
  {{Matthaeus}},\ and\ \citenamefont {{Veltri}}}]{ServidioEA15}%
  \BibitemOpen
  \bibfield  {author} {\bibinfo {author} {\bibfnamefont {S.}~\bibnamefont
  {{Servidio}}}, \bibinfo {author} {\bibfnamefont {F.}~\bibnamefont
  {{Valentini}}}, \bibinfo {author} {\bibfnamefont {D.}~\bibnamefont
  {{Perrone}}}, \bibinfo {author} {\bibfnamefont {A.}~\bibnamefont {{Greco}}},
  \bibinfo {author} {\bibfnamefont {F.}~\bibnamefont {{Califano}}}, \bibinfo
  {author} {\bibfnamefont {W.~H.}\ \bibnamefont {{Matthaeus}}}, \ and\ \bibinfo
  {author} {\bibfnamefont {P.}~\bibnamefont {{Veltri}}},\ }\href {\doibase
  10.1017/S0022377814000841} {\bibfield  {journal} {\bibinfo  {journal}
  {Journal of Plasma Physics}\ }\textbf {\bibinfo {volume} {81}},\ \bibinfo
  {eid} {325810107} (\bibinfo {year} {2015})}\BibitemShut {NoStop}%
\bibitem [{\citenamefont {Valentini}\ \emph {et~al.}(2007)\citenamefont
  {Valentini}, \citenamefont {Tr{\'a}vn{\'\i}{\v{c}}ek}, \citenamefont
  {Califano}, \citenamefont {Hellinger},\ and\ \citenamefont
  {Mangeney}}]{valentini07}%
  \BibitemOpen
  \bibfield  {author} {\bibinfo {author} {\bibfnamefont {F.}~\bibnamefont
  {Valentini}}, \bibinfo {author} {\bibfnamefont {P.}~\bibnamefont
  {Tr{\'a}vn{\'\i}{\v{c}}ek}}, \bibinfo {author} {\bibfnamefont
  {F.}~\bibnamefont {Califano}}, \bibinfo {author} {\bibfnamefont
  {P.}~\bibnamefont {Hellinger}}, \ and\ \bibinfo {author} {\bibfnamefont
  {A.}~\bibnamefont {Mangeney}},\ }\href@noop {} {\bibfield  {journal}
  {\bibinfo  {journal} {Journal of Computational Physics}\ }\textbf {\bibinfo
  {volume} {225}},\ \bibinfo {pages} {753} (\bibinfo {year}
  {2007})}\BibitemShut {NoStop}%
\bibitem [{\citenamefont {Perrone}, \citenamefont {Valentini},\ and\
  \citenamefont {Veltri}(2011)}]{perrone2011role}%
  \BibitemOpen
  \bibfield  {author} {\bibinfo {author} {\bibfnamefont {D.}~\bibnamefont
  {Perrone}}, \bibinfo {author} {\bibfnamefont {F.}~\bibnamefont {Valentini}},
  \ and\ \bibinfo {author} {\bibfnamefont {P.}~\bibnamefont {Veltri}},\ }\href
  {http://stacks.iop.org/0004-637X/741/i=1/a=43} {\bibfield  {journal}
  {\bibinfo  {journal} {The Astrophysical Journal}\ }\textbf {\bibinfo {volume}
  {741}},\ \bibinfo {pages} {43} (\bibinfo {year} {2011})}\BibitemShut
  {NoStop}%
\bibitem [{\citenamefont {Bale}\ \emph {et~al.}(2005)\citenamefont {Bale},
  \citenamefont {Kellogg}, \citenamefont {Mozer}, \citenamefont {Horbury},\
  and\ \citenamefont {Reme}}]{Bale05}%
  \BibitemOpen
  \bibfield  {author} {\bibinfo {author} {\bibfnamefont {S.~D.}\ \bibnamefont
  {Bale}}, \bibinfo {author} {\bibfnamefont {P.~J.}\ \bibnamefont {Kellogg}},
  \bibinfo {author} {\bibfnamefont {F.~S.}\ \bibnamefont {Mozer}}, \bibinfo
  {author} {\bibfnamefont {T.~S.}\ \bibnamefont {Horbury}}, \ and\ \bibinfo
  {author} {\bibfnamefont {H.}~\bibnamefont {Reme}},\ }\href {\doibase
  10.1103/PhysRevLett.94.215002} {\bibfield  {journal} {\bibinfo  {journal}
  {Physical Review Letters}\ }\textbf {\bibinfo {volume} {94}},\ \bibinfo
  {pages} {215002} (\bibinfo {year} {2005})}\BibitemShut {NoStop}%
\bibitem [{Not()}]{NotePezziEA18}%
  \BibitemOpen
  \href@noop {} {\bibinfo  {journal} {Note that, with the present model and
  setup, we describe only the inertial range of turbulence and the
  proton-transition scales. The investigation of sub-proton and electron scale
  turbulence would require a full Vlasov description}\ }\BibitemShut {NoStop}%
\bibitem [{\citenamefont {{Retin{\`o}}}\ \emph {et~al.}(2007)\citenamefont
  {{Retin{\`o}}}, \citenamefont {{Sundkvist}}, \citenamefont {{Vaivads}},
  \citenamefont {{Mozer}}, \citenamefont {{Andr{\'e}}},\ and\ \citenamefont
  {{Owen}}}]{Retino:etal:2007}%
  \BibitemOpen
\bibfield  {journal} {  }\bibfield  {author} {\bibinfo {author} {\bibfnamefont
  {A.}~\bibnamefont {{Retin{\`o}}}}, \bibinfo {author} {\bibfnamefont
  {D.}~\bibnamefont {{Sundkvist}}}, \bibinfo {author} {\bibfnamefont
  {A.}~\bibnamefont {{Vaivads}}}, \bibinfo {author} {\bibfnamefont
  {F.}~\bibnamefont {{Mozer}}}, \bibinfo {author} {\bibfnamefont
  {M.}~\bibnamefont {{Andr{\'e}}}}, \ and\ \bibinfo {author} {\bibfnamefont
  {C.~J.}\ \bibnamefont {{Owen}}},\ }\href {\doibase 10.1038/nphys574}
  {\bibfield  {journal} {\bibinfo  {journal} {Nature Physics}\ }\textbf
  {\bibinfo {volume} {3}},\ \bibinfo {pages} {236} (\bibinfo {year}
  {2007})}\BibitemShut {NoStop}%
\bibitem [{\citenamefont {{Osman}}\ \emph {et~al.}(2011)\citenamefont
  {{Osman}}, \citenamefont {{Matthaeus}}, \citenamefont {{Greco}},\ and\
  \citenamefont {{Servidio}}}]{OsmanEA11}%
  \BibitemOpen
  \bibfield  {author} {\bibinfo {author} {\bibfnamefont {K.~T.}\ \bibnamefont
  {{Osman}}}, \bibinfo {author} {\bibfnamefont {W.~H.}\ \bibnamefont
  {{Matthaeus}}}, \bibinfo {author} {\bibfnamefont {A.}~\bibnamefont
  {{Greco}}}, \ and\ \bibinfo {author} {\bibfnamefont {S.}~\bibnamefont
  {{Servidio}}},\ }\href {\doibase 10.1088/2041-8205/727/1/L11} {\bibfield
  {journal} {\bibinfo  {journal} {Astrophys. J. Lett.}\ }\textbf {\bibinfo
  {volume} {727}},\ \bibinfo {eid} {L11} (\bibinfo {year} {2011})}\BibitemShut
  {NoStop}%
\bibitem [{\citenamefont {{Karimabadi}}\ \emph {et~al.}(2013)\citenamefont
  {{Karimabadi}}, \citenamefont {{Roytershteyn}}, \citenamefont {{Wan}},
  \citenamefont {{Matthaeus}}, \citenamefont {{Daughton}}, \citenamefont
  {{Wu}}, \citenamefont {{Shay}}, \citenamefont {{Loring}}, \citenamefont
  {{Borovsky}}, \citenamefont {{Leonardis}}, \citenamefont {{Chapman}},\ and\
  \citenamefont {{Nakamura}}}]{KarimabadiEA13}%
  \BibitemOpen
  \bibfield  {author} {\bibinfo {author} {\bibfnamefont {H.}~\bibnamefont
  {{Karimabadi}}}, \bibinfo {author} {\bibfnamefont {V.}~\bibnamefont
  {{Roytershteyn}}}, \bibinfo {author} {\bibfnamefont {M.}~\bibnamefont
  {{Wan}}}, \bibinfo {author} {\bibfnamefont {W.~H.}\ \bibnamefont
  {{Matthaeus}}}, \bibinfo {author} {\bibfnamefont {W.}~\bibnamefont
  {{Daughton}}}, \bibinfo {author} {\bibfnamefont {P.}~\bibnamefont {{Wu}}},
  \bibinfo {author} {\bibfnamefont {M.}~\bibnamefont {{Shay}}}, \bibinfo
  {author} {\bibfnamefont {B.}~\bibnamefont {{Loring}}}, \bibinfo {author}
  {\bibfnamefont {J.}~\bibnamefont {{Borovsky}}}, \bibinfo {author}
  {\bibfnamefont {E.}~\bibnamefont {{Leonardis}}}, \bibinfo {author}
  {\bibfnamefont {S.~C.}\ \bibnamefont {{Chapman}}}, \ and\ \bibinfo {author}
  {\bibfnamefont {T.~K.~M.}\ \bibnamefont {{Nakamura}}},\ }\href {\doibase
  10.1063/1.4773205} {\bibfield  {journal} {\bibinfo  {journal} {Physics of
  Plasmas}\ }\textbf {\bibinfo {volume} {20}},\ \bibinfo {eid} {012303}
  (\bibinfo {year} {2013})}\BibitemShut {NoStop}%
\bibitem [{\citenamefont {Drake}\ \emph {et~al.}(2003)\citenamefont {Drake},
  \citenamefont {Swisdak}, \citenamefont {Cattell}, \citenamefont {Shay},
  \citenamefont {Rogers},\ and\ \citenamefont {Zeile}}]{Drake03}%
  \BibitemOpen
  \bibfield  {author} {\bibinfo {author} {\bibfnamefont {F.~M.}\ \bibnamefont
  {Drake}}, \bibinfo {author} {\bibfnamefont {M.}~\bibnamefont {Swisdak}},
  \bibinfo {author} {\bibfnamefont {C.}~\bibnamefont {Cattell}}, \bibinfo
  {author} {\bibfnamefont {M.~A.}\ \bibnamefont {Shay}}, \bibinfo {author}
  {\bibfnamefont {B.~N.}\ \bibnamefont {Rogers}}, \ and\ \bibinfo {author}
  {\bibfnamefont {A.}~\bibnamefont {Zeile}},\ }\href {\doibase
  10.1126/science.1080333} {\bibfield  {journal} {\bibinfo  {journal}
  {Science}\ }\textbf {\bibinfo {volume} {7}},\ \bibinfo {pages} {843}
  (\bibinfo {year} {2003})}\BibitemShut {NoStop}%
\bibitem [{\citenamefont {{Drake}}\ \emph {et~al.}(2010)\citenamefont
  {{Drake}}, \citenamefont {{Opher}}, \citenamefont {{Swisdak}},\ and\
  \citenamefont {{Chamoun}}}]{DrakeEA10}%
  \BibitemOpen
  \bibfield  {author} {\bibinfo {author} {\bibfnamefont {J.~F.}\ \bibnamefont
  {{Drake}}}, \bibinfo {author} {\bibfnamefont {M.}~\bibnamefont {{Opher}}},
  \bibinfo {author} {\bibfnamefont {M.}~\bibnamefont {{Swisdak}}}, \ and\
  \bibinfo {author} {\bibfnamefont {J.~N.}\ \bibnamefont {{Chamoun}}},\ }\href
  {\doibase 10.1088/0004-637X/709/2/963} {\bibfield  {journal} {\bibinfo
  {journal} {The Astrophysical Journal}\ }\textbf {\bibinfo {volume} {709}},\
  \bibinfo {pages} {963} (\bibinfo {year} {2010})},\ \Eprint
  {http://arxiv.org/abs/0911.3098} {arXiv:0911.3098 [astro-ph.SR]} \BibitemShut
  {NoStop}%
\bibitem [{\citenamefont {Swisdak}(2016)}]{swisdak2016quantifying}%
  \BibitemOpen
  \bibfield  {author} {\bibinfo {author} {\bibfnamefont {M.}~\bibnamefont
  {Swisdak}},\ }\href@noop {} {\bibfield  {journal} {\bibinfo  {journal}
  {Geophysical Research Letters}\ }\textbf {\bibinfo {volume} {43}},\ \bibinfo
  {pages} {43} (\bibinfo {year} {2016})}\BibitemShut {NoStop}%
\bibitem [{\citenamefont {Pezzi}\ \emph
  {et~al.}(2017{\natexlab{c}})\citenamefont {Pezzi}, \citenamefont {Parashar},
  \citenamefont {Servidio}, \citenamefont {Valentini}, \citenamefont
  {V\'asconez}, \citenamefont {Yang}, \citenamefont {Malara}, \citenamefont
  {Matthaeus},\ and\ \citenamefont {Veltri}}]{pezzi2017colliding}%
  \BibitemOpen
  \bibfield  {author} {\bibinfo {author} {\bibfnamefont {O.}~\bibnamefont
  {Pezzi}}, \bibinfo {author} {\bibfnamefont {T.~N.}\ \bibnamefont {Parashar}},
  \bibinfo {author} {\bibfnamefont {S.}~\bibnamefont {Servidio}}, \bibinfo
  {author} {\bibfnamefont {F.}~\bibnamefont {Valentini}}, \bibinfo {author}
  {\bibfnamefont {C.~L.}\ \bibnamefont {V\'asconez}}, \bibinfo {author}
  {\bibfnamefont {Y.}~\bibnamefont {Yang}}, \bibinfo {author} {\bibfnamefont
  {F.}~\bibnamefont {Malara}}, \bibinfo {author} {\bibfnamefont {W.~H.}\
  \bibnamefont {Matthaeus}}, \ and\ \bibinfo {author} {\bibfnamefont
  {P.}~\bibnamefont {Veltri}},\ }\href {\doibase 10.1017/S0022377817000113}
  {\bibfield  {journal} {\bibinfo  {journal} {Journal of Plasma Physics}\
  }\textbf {\bibinfo {volume} {83}},\ \bibinfo {pages} {705830108} (\bibinfo
  {year} {2017}{\natexlab{c}})}\BibitemShut {NoStop}%
\bibitem [{\citenamefont {Grad}(1949)}]{Grad49}%
  \BibitemOpen
  \bibfield  {author} {\bibinfo {author} {\bibfnamefont {H.}~\bibnamefont
  {Grad}},\ }\href@noop {} {\bibfield  {journal} {\bibinfo  {journal} {Commun.
  Pure Appl. Math.}\ }\textbf {\bibinfo {volume} {2}},\ \bibinfo {pages} {331}
  (\bibinfo {year} {1949})}\BibitemShut {NoStop}%
\bibitem [{\citenamefont {{Armstrong}}\ and\ \citenamefont
  {{Montgomery}}(1967)}]{ArmstrongMontgomery67}%
  \BibitemOpen
  \bibfield  {author} {\bibinfo {author} {\bibfnamefont {T.}~\bibnamefont
  {{Armstrong}}}\ and\ \bibinfo {author} {\bibfnamefont {D.}~\bibnamefont
  {{Montgomery}}},\ }\href {\doibase 10.1017/S0022377800003421} {\bibfield
  {journal} {\bibinfo  {journal} {Journal of Plasma Physics}\ }\textbf
  {\bibinfo {volume} {1}},\ \bibinfo {pages} {425} (\bibinfo {year}
  {1967})}\BibitemShut {NoStop}%
\bibitem [{\citenamefont {{Schumer}}\ and\ \citenamefont
  {{Holloway}}(1998)}]{SchumerHolloway98}%
  \BibitemOpen
  \bibfield  {author} {\bibinfo {author} {\bibfnamefont {J.~W.}\ \bibnamefont
  {{Schumer}}}\ and\ \bibinfo {author} {\bibfnamefont {J.~P.}\ \bibnamefont
  {{Holloway}}},\ }\href {\doibase 10.1006/jcph.1998.5925} {\bibfield
  {journal} {\bibinfo  {journal} {Journal of Computational Physics}\ }\textbf
  {\bibinfo {volume} {144}},\ \bibinfo {pages} {626} (\bibinfo {year}
  {1998})}\BibitemShut {NoStop}%
\bibitem [{\citenamefont {Yin}(2014)}]{Zhaohua14}%
  \BibitemOpen
  \bibfield  {author} {\bibinfo {author} {\bibfnamefont {Z.}~\bibnamefont
  {Yin}},\ }\href {\doibase https://doi.org/10.1016/j.jcp.2013.10.039}
  {\bibfield  {journal} {\bibinfo  {journal} {Journal of Computational
  Physics}\ }\textbf {\bibinfo {volume} {258}},\ \bibinfo {pages} {371 }
  (\bibinfo {year} {2014})}\BibitemShut {NoStop}%
\bibitem [{\citenamefont {{Knorr}}(1977)}]{Knorr77}%
  \BibitemOpen
  \bibfield  {author} {\bibinfo {author} {\bibfnamefont {G.}~\bibnamefont
  {{Knorr}}},\ }\href {\doibase 10.1088/0032-1028/19/6/004} {\bibfield
  {journal} {\bibinfo  {journal} {Plasma Physics}\ }\textbf {\bibinfo {volume}
  {19}},\ \bibinfo {pages} {529} (\bibinfo {year} {1977})}\BibitemShut
  {NoStop}%
\bibitem [{\citenamefont {Kennel}\ and\ \citenamefont
  {Petschek}(1966)}]{Kennel66}%
  \BibitemOpen
  \bibfield  {author} {\bibinfo {author} {\bibfnamefont {C.~F.}\ \bibnamefont
  {Kennel}}\ and\ \bibinfo {author} {\bibfnamefont {H.}~\bibnamefont
  {Petschek}},\ }\href@noop {} {\bibfield  {journal} {\bibinfo  {journal}
  {Journal of Geophysical Research}\ }\textbf {\bibinfo {volume} {71}},\
  \bibinfo {pages} {1} (\bibinfo {year} {1966})}\BibitemShut {NoStop}%
\bibitem [{\citenamefont {Shebalin}, \citenamefont {Matthaeus},\ and\
  \citenamefont {Montgomery}(1983)}]{Shebalin83}%
  \BibitemOpen
  \bibfield  {author} {\bibinfo {author} {\bibfnamefont {J.~V.}\ \bibnamefont
  {Shebalin}}, \bibinfo {author} {\bibfnamefont {W.~H.}\ \bibnamefont
  {Matthaeus}}, \ and\ \bibinfo {author} {\bibfnamefont {D.}~\bibnamefont
  {Montgomery}},\ }\href {\doibase 10.1017/S0022377800000933} {\bibfield
  {journal} {\bibinfo  {journal} {J. Plasma Phys.}\ }\textbf {\bibinfo {volume}
  {29}},\ \bibinfo {pages} {525} (\bibinfo {year} {1983})}\BibitemShut
  {NoStop}%
\bibitem [{\citenamefont {{Sircombe}}, \citenamefont {{Arber}},\ and\
  \citenamefont {{Dendy}}(2006)}]{SircombeEA06}%
  \BibitemOpen
  \bibfield  {author} {\bibinfo {author} {\bibfnamefont {N.~J.}\ \bibnamefont
  {{Sircombe}}}, \bibinfo {author} {\bibfnamefont {T.~D.}\ \bibnamefont
  {{Arber}}}, \ and\ \bibinfo {author} {\bibfnamefont {R.~O.}\ \bibnamefont
  {{Dendy}}},\ }\href {\doibase 10.1051/jp4:2006133055} {\bibfield  {journal}
  {\bibinfo  {journal} {Journal de Physique IV}\ }\textbf {\bibinfo {volume}
  {133}},\ \bibinfo {pages} {277} (\bibinfo {year} {2006})}\BibitemShut
  {NoStop}%
\bibitem [{\citenamefont {{Schekochihin}}\ \emph {et~al.}(2008)\citenamefont
  {{Schekochihin}}, \citenamefont {{Cowley}}, \citenamefont {{Dorland}},
  \citenamefont {{Hammett}}, \citenamefont {{Howes}}, \citenamefont {{Plunk}},
  \citenamefont {{Quataert}},\ and\ \citenamefont
  {{Tatsuno}}}]{SchekochihinEA08}%
  \BibitemOpen
  \bibfield  {author} {\bibinfo {author} {\bibfnamefont {A.~A.}\ \bibnamefont
  {{Schekochihin}}}, \bibinfo {author} {\bibfnamefont {S.~C.}\ \bibnamefont
  {{Cowley}}}, \bibinfo {author} {\bibfnamefont {W.}~\bibnamefont {{Dorland}}},
  \bibinfo {author} {\bibfnamefont {G.~W.}\ \bibnamefont {{Hammett}}}, \bibinfo
  {author} {\bibfnamefont {G.~G.}\ \bibnamefont {{Howes}}}, \bibinfo {author}
  {\bibfnamefont {G.~G.}\ \bibnamefont {{Plunk}}}, \bibinfo {author}
  {\bibfnamefont {E.}~\bibnamefont {{Quataert}}}, \ and\ \bibinfo {author}
  {\bibfnamefont {T.}~\bibnamefont {{Tatsuno}}},\ }\href {\doibase
  10.1088/0741-3335/50/12/124024} {\bibfield  {journal} {\bibinfo  {journal}
  {Plasma Physics and Controlled Fusion}\ }\textbf {\bibinfo {volume} {50}},\
  \bibinfo {eid} {124024} (\bibinfo {year} {2008})},\ \Eprint
  {http://arxiv.org/abs/0806.1069} {arXiv:0806.1069 [physics.plasm-ph]}
  \BibitemShut {NoStop}%
\bibitem [{\citenamefont {Hatch}\ \emph {et~al.}(2014)\citenamefont {Hatch},
  \citenamefont {Jenko}, \citenamefont {Bratanov},\ and\ \citenamefont
  {Navarro}}]{HatchEA14}%
  \BibitemOpen
  \bibfield  {author} {\bibinfo {author} {\bibfnamefont {D.~R.}\ \bibnamefont
  {Hatch}}, \bibinfo {author} {\bibfnamefont {F.}~\bibnamefont {Jenko}},
  \bibinfo {author} {\bibfnamefont {V.}~\bibnamefont {Bratanov}}, \ and\
  \bibinfo {author} {\bibfnamefont {A.~B.}\ \bibnamefont {Navarro}},\ }\href
  {\doibase 10.1017/S0022377814000154} {\bibfield  {journal} {\bibinfo
  {journal} {Journal of Plasma Physics}\ }\textbf {\bibinfo {volume} {80}},\
  \bibinfo {pages} {531–551} (\bibinfo {year} {2014})}\BibitemShut {NoStop}%
\bibitem [{\citenamefont {{Parker}}\ and\ \citenamefont
  {{Dellar}}(2015)}]{ParkerDellar15}%
  \BibitemOpen
  \bibfield  {author} {\bibinfo {author} {\bibfnamefont {J.~T.}\ \bibnamefont
  {{Parker}}}\ and\ \bibinfo {author} {\bibfnamefont {P.~J.}\ \bibnamefont
  {{Dellar}}},\ }\href {\doibase 10.1017/S0022377814001287} {\bibfield
  {journal} {\bibinfo  {journal} {Journal of Plasma Physics}\ }\textbf
  {\bibinfo {volume} {81}},\ \bibinfo {eid} {305810203} (\bibinfo {year}
  {2015})},\ \Eprint {http://arxiv.org/abs/1407.1932} {arXiv:1407.1932
  [physics.plasm-ph]} \BibitemShut {NoStop}%
\bibitem [{\citenamefont {{Adkins}}\ and\ \citenamefont
  {{Schekochihin}}(2018)}]{Adkins18}%
  \BibitemOpen
  \bibfield  {author} {\bibinfo {author} {\bibfnamefont {T.}~\bibnamefont
  {{Adkins}}}\ and\ \bibinfo {author} {\bibfnamefont {A.~A.}\ \bibnamefont
  {{Schekochihin}}},\ }\href {\doibase 10.1017/S0022377818000089} {\bibfield
  {journal} {\bibinfo  {journal} {Journal of Plasma Physics}\ }\textbf
  {\bibinfo {volume} {84}},\ \bibinfo {pages} {905840107} (\bibinfo {year}
  {2018})}\BibitemShut {NoStop}%
\end{thebibliography}

%

\end{document}